\documentclass[twocolumn,secnumarabic,amssymb, nobibnotes, aps, prd]{revtex4}
\usepackage{graphicx}

\begin{document}
\title{The fundamental obscurity in quantum mechanics. \\ Could the problem be considered universal?}
\author{A. V. Nikulov}
%\email[]{nikulov@ipmt-hpm.ac.ru}
\affiliation{Institute of Microelectronics Technology and High Purity Materials, Russian Academy of Sciences, 142432 Chernogolovka, Moscow District, RUSSIA.} %nikulov@ipmt-hpm.ac.ru
%\date{}
\begin{abstract}The contemporary controversy about the fundamental obscurity in quantum mechanics keeps on the old one about the aim of science, which was between the founders of the quantum theory. The orthodox quantum mechanics could be created only at the cost of renunciation of  reality as the aim of natural science. The description only of phenomena, i.e. results of observation, should not be universal if no one believes that these phenomena are manifestation of a unique reality. Such belief concerning quantum mechanics is quite unacceptable because of irremediably conflict with special relativity. Nevertheless the quantum mechanics was developed and apprehended by most physicists as a universal theory of a quantum world. This fundamental discrepancy between the essence of the orthodox quantum mechanics and its history of development and studying has resulted both to an illusion about the aim of its description among most physicists and to the consideration of its fundamental obscurity as a universal problem among experts in quantum foundation. The aim of this paper is to show that quantum phenomena can not be described universally. It is indicated that rather the Schrodinger's than Born's interpretation of the wave function is valid for description of many quantum phenomena. The fundamental obscurity with which we are faced at the description, for example, macroscopic quantum phenomena differs fundamentally from the one with which the founders of the quantum theory were faced on atomic level.
 \end{abstract}

\maketitle

\narrowtext

\section*{Introduction}
%\label{intro}
The duality is one of the most paradoxical features of quantum phenomena. And not only. The duality is also the distinctive feature of the attitude to quantum mechanics and of its history. On the one hand, quantum mechanics (QM) is generally regarded by most physicists as the physical theory that is our best candidate for a fundamental and universal description of the physical world. But on the other hand some experts understand {\it that QM is still a not-yet-completely-understood theory open to further fundamental research} \cite{Nikolic2007}. This dual attitude may be considered as a result of lack of correspondence between the essence of QM and the history of its development. The quantum theory was emerged in order to describe the paradoxical phenomena with which physicists were faced on atomic level. The formal rules of quantization, proposed by Plank, Einstein and Bohr at the beginning of the XX century, allowed to describe some of these phenomena. But as Heisenberg noted in 1925 \cite{Heisenberg1925} these formal rules provoked serious objections. These rules included relations between variables, which can not be observed, such as position and circulation time of electron on atomic orbit. Heisenberg proposed to create a QM considering relations only between observable variables \cite{Heisenberg1925}. This proposal to do not consider {\it hidden variables} results to creation of the QM studied last eighty years. 

Most breakthroughs of the XX century physics are fairly connected with this QM. But the revision of the aim of physics proposed by Heisenberg has provoked the disagreement and the controversy of many years between the founding fathers of quantum theory. This epistemological controversy was just about the aim of science. Einstein {\it fully recognised the very important progress which the statistical quantum theory has brought to theoretical physics} \cite{Einstein1949}. The {\it reasons which keep he from falling in line with the opinion of almost all contemporary theoretical physicists} \cite{Einstein1949} were explained quite clearly in his utterances, for example in \cite{Einstein1949}: {\it "What does not satisfy me in that theory, from the standpoint of principle, is its attitude towards that which appears to me to be the programmatic aim of all physics: the complete description of any (individual) real situation (as it supposedly exists irrespective of any act of observation or substantiation)"}. Einstein described also quite clearly the philosophical beliefs of his opponents: {\it "Whenever the positivistically inclined modern physicist hears such a formulation his reaction is that of a pitying smile. He says to himself: "there we have the naked formulation of a metaphysical prejudice, empty of content, a prejudice, moreover, the conquest of which constitutes the major epistemological achievement of physicists within the last quarter-century. Has any man ever perceived a 'real physical situation'? How is it possible that a reasonable person could today still believe that he can refute our essential knowledge and understanding by drawing up such a bloodless ghost?""} \cite{Einstein1949}. 

The creators of the QM defended the completeness of its description on base of the positivistic beliefs, according to which the theory should describe no real processes but only results of observations: {\it "A real difficulty in the understanding of the Copenhagen interpretation arises, however, when one asks the famous question: But what happens 'really' in an atomic event?"} \cite{Heisenberg1958}; {\it "There is no quantum world. There is only an abstract quantum physical description. It is wrong to think that the task of physics is to find out how Nature is"} \cite{Bohr1958}. Heisenberg and Bohr have lost hope to understand {\it what happens between two observations} because, as Heisenberg stated \cite{Heisenberg1958}, {\it "Any attempt to find such a description would lead to contradictions"} and therefore {\it "the term 'happens' is restricted to the observation"}. But the QM generates puzzles even in the limits of this pure positivism. Some of these puzzles were intelligible right from the start of the orthodox QM. Bohr wrote in \cite{Bohr1949} that at the Solvay meeting 1928 "{\it an interesting discussion arose also about how to speak of the appearance of phenomena for which only predictions of statistical character can be made. The question was whether, as to the occurrence of individual effects, we should adopt a terminology proposed by Dirac, that we were concerned with a choice on the part of "nature" or, as suggested by Heisenberg, we should say that we have to do with a choice on the part of the "observer" constructing the measuring instruments and reading their recording. Any such terminology would, however, appear dubious since, on the one hand, it is hardly reasonable to endow nature with volition in the ordinary sense, while, on the other hand, it is certainly not possible for the observer to influence the events which may appear under the conditions he has arranged}". Heisenberg insisted {\it on the part of the "observer"} during a long time: {\it "In classical physics science started from the belief - or should one say from the illusion? - that we could describe the world or at least parts of the world without any reference to ourselves"} \cite{Heisenberg1958}. But he had not proposed how the world could be exactly described {\it with a reference to ourselves}. A continuation of the discussion at the Solvay meeting 1928 about a  choice of measurement result may be seen in some contemporary debate, for example about "free will" \cite{Hooft2007}. 
	
Two famous puzzles generated by QM were revealed by Einstein and Schrodinger. The both puzzles cast doubt on the cardinal positive principle of the orthodox QM (according for example the textbook \cite{LandauL}), superposition of states. Einstein raised the problem of a non-locality of the QM proposed by Heisenberg and Bohr as far back as 1927 \cite{Einstein1927} and pointed out first in this talk at the 5th Solvay meeting that the interpretation of the Schrodinger wave function as a description of an individual particle motion results to contradiction with the relativity principle. The non-locality was revealed in a more subtle form in the famous paper by Einstein, Podolsky and Rosen \cite{EPR1935}. Motivated by the EPR paper Schrodinger coined the term "entanglement" \cite{Schrod35D,Schrod35E}. The essence of entanglement \cite{Entangl} (EPR correlation) has provoked the controversy of many years among experts in foundations of QM which is in progress up to now \cite{Norsen09,Ghirardi10}. 

The outstanding contribution to this debate was made by John Bell. The EPR paradox \cite{EPR1935} has revealed the non-locality of QM in description which may depend on an interpretation of this description. It is called by John Cramer {\it non-locality of the first kind} \cite{Cramer1986}. The contradiction of the non-locality of the first kind with the relativity principle can be easy eliminated with help of the interpretation that the superposition of states (the Schrodinger wave function, or the state vector) represents our knowledge of the system we are trying to describe. The authors \cite{Entangl} accentuate that Schrodinger coined the term {\it "entanglement of {\bf our knowledge}"} to describe the situation considered by EPR \cite{EPR1935}. The identification of the state vector with "knowledge of the system" by Heisenberg is defined in \cite{Cramer1986} as the fourth principal element of the Copenhagen interpretation, which eliminates simple non-locality problems. The famous work by John Bell \cite{Bell1964} has allowed to reveal the non-locality on a more deeper level, at observations. Interpretations become irrelevant because real observations are involved in this {\it non-locality of the second kind} \cite{Cramer1986}. The importance of the Bell's theorem was perceived during many years even by experts in foundations of QM \cite{Ballantine87}, see also p.125 in the book \cite{QuCh2006}. Now many physicists are aware on the Bell's theorem and almost all physicists heard at last on so-called Bell Inequalities. But interpretations of the no-hidden-variables theorem of Bell are not merely dual. There may be seen at least two levels of duality. The first level of the duality was designated by Mermin as far back as 1985 \cite{Mermin1985}: {\it In the question of whether there is some fundamental problem with QM signaled by tests of Bell's inequality, physicists can be divided into a majority who are "indifferent" and a minority who are "bothered".} The indifference of the majority can be observed in the most modern publications. For example, the authors of the book \cite{Chuang2000} and many other publications about quantum computation are sure that violation of the Bell's inequalities proves only that Einstein was not right and that the orthodox QM is correct theory. The second level of the duality may be seen among the minority who are "bothered". This duality is provoked first of all with the question: "Does quantum non-locality irremediably conflict with special relativity?" \cite{Ghirardi10}.

This question results directly from the EPR paradox \cite{EPR1935}. States of two particles are entangled with help of a conservation law in \cite{EPR1935}. Schrodinger entangling states of atom and cat with experimental conditions \cite{Schrod35D} has manifested other famous puzzle connected with the non-determinism of QM. The duality is typical for the attitude also to this "Schrodinger's cat" paradox. Since all cats, which we know, are macroscopic some authors \cite{QuCh2006,cat1,cat2} associate this paradox with problems of macroscopic quantum phenomena. But anyone should easy understand that nothing in the Schrodinger paradox could depend on size of the cat. Moreover, anyone should easy understand that nothing, except tragedy situation, could change in this paradox at substitution of cat, small flask of hydrocyanic acid and hammer for a recorder, which can record the discharge of Geiger counter tube. Therefore the Schrodinger cat paradox should associate with a fuzzy status of measurements in QM \cite{Bell1987,Bell1990}, discussed hotly among the minority who are "bothered" \cite{Mermin2006,Ghirardi2008,Mermin2008}.

The absence of unity concerning QM indicates its fundamental obscurity noted by Bell \cite{Bell1984} and other experts. The diversity of the numerous interpretations proposed during the QM history witness that {\it quantum mechanics is not yet based on a generally accepted conceptual foundation} \cite{Zeil1999}. Nevertheless almost full unity of views is observed concerning an aspect. According to the common belief QM should and can be considered as a universal theory. The belief in the universality of most physicists is not conscious and only few authors claim this universality explicitly \cite{Kiefer2010}. The belief in the universality of a physical theory is implicitly based on a confidence in existence of a unique reality, which the theory should describe. The author \cite{Kiefer2010} realises this connection between the belief in the universality of the quantum theory and the quantum reality and confesses that {\it "the consequences of this universality for our world view seem enormous, since the exact linearity of the theory would directly lead, in a realist interpretation, to an Everett-type of interpretation. Such a view is disapproved by many physicists because of its paradoxical touch"}. The {\it many physicists} approve the orthodox QM which was created on the base of repudiation of a reality as the aim of description. The description of phenomena (i.e. results of observations) should not demand universality. Nevertheless the founding fathers of QM did not relinquish the demand of universality in defiance of logic and all physicists do not call this universality in question. A duality is even in this unity. The minority who are "bothered" consider the fundamental obscurity in QM as a universal problem whereas the majority who are "indifferent" are sure that QM is universal description of the physical world. In this paper I try to show that the fundamental obscurity in QM can not be considered universally and that the baseless belief in the universality of quantum description results to actual mistakes. It will be shown that the fundamental obscurity at the description of macroscopic quantum phenomena differs fundamentally from the one associated with the EPR correlation and fuzzy concept of measurement in QM.  

\section{The essence of the fundamental obscurity discussed now}
%\label{sec:2}
Bell connected the fundamental obscurity in QM with the Problem: {\it "how exactly is the world to be divided into speakable apparatus...that we can talk about...and unspeakable quantum system that we can not talk about?"} \cite{Bell1984}. He explained \cite{Bell1984} why this Problem could emerge: {\it "The founding fathers of quantum theory decided even that no concepts could possibly be found which could permit direct description of the quantum world. So the theory which they established aimed only to describe systematically the response of the apparatus."} Bohr, Heisenberg and others contended that this direct description is not indispensable because of {\it "the impossibility of any sharp separation between the behaviour of atomic objects and the interaction with the measuring instruments which serve to define the conditions under which the phenomena appear"} \cite{Bohr1949}. According to the quantum postulate proposed by Bohr in 1928 {\it any observation of atomic phenomena should include an interaction they with equipment used for the observation which can not be neglected} \cite{Bohr1928}. The uncertainty principle proposed by Heisenberg in 1927 \cite{Heisenberg1927} postulates that it is impossible to measure simultaneously some variables (canonically conjugate quantities), for example position and momentum of a particle, with any great degree of accuracy or certainty. And the complementarity principle proposed by Bohr should explain why it is impossible. Bohr stressed that {\it "an adequate tool for a complementary way of description is offered precisely by the quantum-mechanical formalism"} \cite{Bohr1949}. This formalism turned out very successful in describing quantum phenomena, but no a quantum reality. Most physicists, as Karl Popper noted \cite{Popper}, have not comprehended that the Copenhagen interpretation should imply the renunciation of realism. In fact the QM developed and was studied as a description of a quantum world in spite of the Bohr's statement that {\it There is no quantum world}.  

Because of this duality of the QM history it is important to note that the phenomena observed only on atomic level are discussed in the controversy about the renunciation of realism. Bohr discussed with Einstein  epistemological problems in {\bf atomic physics} \cite{Bohr1949}. For example, it is written in his reply \cite{Bohr1935} on the EPR paper \cite{EPR1935}: {\it "The trend of their argumentation, however, does not seem to me adequately to meet the actual situation with which we are faced in {\bf atomic physics}"}. Heisenberg connected {\it A real difficulty in the understanding of the Copenhagen interpretation} with {\it the famous question} about {\it an atomic event} \cite{Heisenberg1958}, see the quote above. Einstein was sure that {\it in the macroscopic sphere it simply is considered certain that one must adhere to the program of a realistic description in space and time} \cite{Einstein1949}. Nevertheless {\it the quantum-mechanical formalism offered for a complementary way of description} \cite{Bohr1949} was formally expanded on macroscopic level at the description of macroscopic quantum phenomena, superfluidity and superconductivity. The fundamental obscurity in QM considered by Bell connected with {\it the inanimate apparatus which amplifies microscopic events to macroscopic consequences} \cite{Bell1984}. But macroscopic phenomena should not be amplified. This obvious fact casts doubt on the universality of this fundamental obscurity.

\subsection{Fuzzy concept of measurement in quantum mechanics.}
%\label{sec:2.1}
The majority of paradoxes and debates \cite{Mermin2006,Ghirardi2008,Mermin2008} are connected with the problem of measurements. The concept of measurement must have fundamental importance for QM describing phenomena, i.e. results of measurements. But as Bell noted justly: {\it "The concept of 'measurement' becomes so fuzzy on reflection that it is quite surprising to have it appearing in physical theory at the most fundamental level"} \cite{Bell1987}. In fact this concept is reduced to the words on the collapse \cite{Neumann1932} of wave function or a 'quantum jump' {\it "from the 'possible' to the 'actual' taking place during the act of observation"} \cite{Heisenberg1958}. The concept of the collapse is the source of the problem of non-locality and of many of the most severe interpretation problems \cite{Cramer1986}. The description of quantum phenomena with the orthodox QM forbids to assume a duration of the collapse in time \cite{QuCh2006}. The collapse must take place instantaneously over all space. To be more precise it must take place outside the physical time and space. Such 'quantum jump' is possible only for our knowledge, which should change when we will have known on results of measurements. The cardinal positive principle of the QM, superposition of states \cite{LandauL} is inconceivable without the collapse because of the obvious impossibility to see anything in several places or to measure different values of a parameter simultaneously. Therefore the superposition of states can describe only our knowledge. This knowledge about two parts of a system can be entangled, so that {\it "Maximal knowledge of a total system does not necessarily include total knowledge of all its parts, not even when these are fully separated from each other and at the moment are not influencing each other at all"} \cite{Schrod35D}. 

\subsection{Could the informational interpretation overcome the fundamental obscurity in quantum mechanics?}
%\label{sec:2.2}
Ironically, the entanglement (EPR correlation) introduced \cite{EPR1935,Schrod35D,Schrod35E} by the opponent of the Copenhagen interpretation in order to prove the incompleteness of quantum description of physical reality has become a basis of the idea of quantum computation. Because of the interest in quantum computation the information interpretation of QM has become popular more than even before. GianCarlo Ghirardi calls this position {\it "the newest orthodoxy: quantum mechanics is not about the world, it is exclusively about information"} \cite{Ghirardi2008}. He reminds \cite{Ghirardi2008} that Bell {\it "refused to consider such a position unless \cite{Bell1990}, in advance, one would have answered to two basic (for him) questions: whose information?, and: information about what?"} The argument between modern experts \cite{Ghirardi2008,Mermin2008} about the information interpretation is no less heated and dramatic than it was between the founding fathers of quantum theory about realism and positivism. The brilliant physicist, David Mermin confessed in 2001 \cite{Mermin2001}: {\it "Until quite recently I was entirely on Bell's side on the matter of knowledge-information. But then I fell into bad company"}. Because of his associations with quantum computer scientists he has {\it "come to feel that "Information about what?" is a fundamentally metaphysical question that ought not to distract tough-minded physicists"} \cite{Mermin2001}. Such radical revision of the attitude to physical theory because of {\it the quantum computation revolution} seems ungrounded. Peter Shor's factoring algorithm or Lov Grover's search algorithm are pure speculative science whereas physics is rather empirical than speculative science. The disparaging attitude to the question "Information about what?" as fundamentally metaphysical \cite{Mermin2001} repeats the one by {\it the positivistically inclined modern physicist} to {\it a real physical situation} \cite{Einstein1949} (see the quote above). The newest orthodoxy seems to do not differ from the old orthodoxy by Heisenberg and Bohr. It is no mere chance that David Mermin has {\it learned to stop worrying and love Bohr} \cite{Mermin2003} when he {\it started hanging out with the quantum computation crowd, for many of whom quantum mechanics is self-evidently and unproblematically all about information} \cite{Mermin2001}. 

David Mermin expresses heartfelt regret \cite{Mermin2001} that we are deprived the possibility to know about Einstein's reaction to the Bell's theorem and Bell's reaction to {\it the quantum computation revolution of the 1990's} \cite{Mermin2001}. Indeed, it is very great loss. But I think that both Einstein and Bell have expound enough clear their opinion about quantum mechanics in order one can conjecture their reaction to the idea of the quantum computation. I assume that Einstein could be surprised that the EPR correlation, which he considered as absolutely unreal, would lay down the foundations of the idea of a real device. Or quantum computer is not a real device!? I assume also that not only Einstein and Bell, but even Heisenberg could be very surprised the statement {\it "that the Copenhagen interpretation should provide a congenial setting for the exposition of quantum computation"} \cite{Mermin2003}. Heisenberg forewarned many times that {\it "there is no way of describing what happens between two consecutive observations"} and {\it "that the concept of the probability function does not allow a description of what happens between two observations"} \cite{Heisenberg1958}. It is well known that the quantum computation should be just between observations. Thus, according to Heisenberg, the QM in its Copenhagen interpretation can not describe a process of quantum computation. David Mermin states that a measurement problem is absent in quantum computer science \cite{Mermin2008}. Indeed, no problem when {\it you take the theory seriously as a source of impossibly fast algorithms for the processing of knowledge-information} \cite{Mermin2001} and limit oneself the speculation about quantum computation. The puzzle emerges when you try to understand what quantum systems can be used for a creation of a real quantum computer. According to Bell it is puzzle because quantum systems are unspeakable {\it that we can not talk about} \cite{Bell1984}.

\subsection{Beables and observables.}
%\label{sec:2.3}
But it is no puzzle for authors of numerous publications which are sure that almost all two-states quantum systems, even macroscopic one \cite{Mooij99,Makhlin01,Leggett02,Mooij03,Clark08}, can be used as quantum bits. Moreover some authors contrive to entangle these "qubits" with help of a real electromagnetic interaction \cite{Averin98,Makhlin99,Berkley03,Ilichev04,Mooij07,Devoret10} in defiance of the intrinsic contradiction of the EPR correlation with local realism \cite{Aristov04,Aristov05}. Bell expressed regret that {\it "few theoretical physicists wanted to hear about"} the de Broglie-Bohm theory \cite{Bell1984}. He said in 1984: {\it "Even now the de Broglie-Bohm picture is generally ignored, and not taught to students. I think this is a great loss. For that picture exercises the mind in a very salutary way} \cite{Bell1984}. This picture is generally ignored up to now. I think that the preposterous idea \cite{Averin98,Makhlin99} to create the EPR correlation with help of a real interaction is one of unfortunate results of this ignorance. It should be clear for anyone who read \cite{Bohm1952} that the entanglement, described realistically by Bohm with help of non-local quantum potential, is possible only if "qubits" must be describe with a common $\psi $ - function. 

The authors \cite{Averin98,Makhlin99,Berkley03,Ilichev04,Mooij07,Devoret10} and other representatives of {\it the quantum computation crowd} \cite{Mermin2001} disregard the intrinsic contradiction of the idea of a quantum computer as a real device. The personal contact with some of these authors has show me that they are sure that the quantum computer is a real device because of the great number of real quantum devices and equipment created in the XX century. But between the quantum computer and all real quantum devices is a fundamental difference. The operation of the real quantum devices created in the XX century does not contradict inevitably to realism as well as most quantum phenomena. Bell accentuated in his  paper \cite{Bell1966} that {\it "It was not the objective measurable predictions of quantum mechanics which ruled out hidden variable"}. The quantum computer can not be possible without the EPR correlation (entanglement) contradicting to local realism inevitably and because of its essence. The inevitable contradiction between {\it the objective measurable predictions} giving the orthodox QM and any realistic description is revealed with help of so called no-hidden-variables theorem (or, vulgarly, no-go theorem) \cite{Mermin1993}. The famous no-hidden-variables theorem by Bell \cite{Bell1964} has revealed the contradiction between {\it the objective measurable predictions} of the EPR correlation and any local realistic description. 

The experimental evidences (for example \cite{Aspect1981}) of violation of the Bell's inequalities testify to the observation of the EPR correlation and against local realism. But there is no reason for the confidence in a real existence of the EPR correlation prevailing among the {\it majority who are "indifferent"}. For Bell the violation of the Bell's inequality was {\it "the real problem with quantum theory: the apparently essential conflict between any sharp formulation and fundamental relativity"} \cite{Bell1984}. The representatives of the {\it minority who are "bothered"} \cite{Norsen09,Ghirardi10,Mermin1998,Zeil2005} dispute about a possibility to get rid of this troubling conflict with the relativistic causality at the cost of renunciation of realism and determinism. Whereas representatives of the {\it majority who are "indifferent"} \cite{Martinis2009,Korotkov2010} extend thoughtlessly and groundlessly this conflict to macroscopic level. 

These two levels of the duality concerning the observation of the EPR correlation may be connected with different attitude to the problem of reality as {\it the programmatic aim of all physics}. Therefore it is needed to define more exactly what is reality. George Berkeley (1685 -1753), bishop of Cloyne, and other philosophers proved that theories of natural philosophy, in particular Newtonian machinery, should not lay a claim to a description of a reality. The theories should be used only as a computational instrument to describe phenomena, because of the obvious impossibility to receive evidence that the phenomena are manifestation of an objective reality. In spite of Berkeley most scientists believed that science should investigate a reality and interpreted without a moment's hesitation all physical phenomena as manifestation of a reality. But some quantum phenomena could not be interpreted in that way. Therefore Heisenberg, Bohr and other adherents of the Copenhagen interpretation have followed to Berkeley. 

Arguing against this deviation from the tradition ascending from Galilei \cite{Popper} Einstein did not state that {\it any man can perceive a 'real physical situation'}. He criticised the adherence of {\it the positivistically inclined modern physicist} to the {\it Berkeley's principle, esse est percipi} \cite{Einstein1949} because of other reason. According his point of view \cite{Einstein1949}: {\it ""Being" is always something which is mentally constructed by us, that is, something which we freely posit (in the logical sense). The justification of such constructs does not lie in their derivation from what is given by the senses. . . . The justification of the constructs, which represent "reality" for us, lies alone in their quality of making intelligible what is sensorily given"}. Explaining further this point of view Einstein writes in \cite{Einstein1949}: {\it "After what has been said, the "real" in physics is to be taken as a type of program, to which we are, however, not forced to cling a priori. No one is likely to be inclined to attempt to give up this program within the realm of the "macroscopic"}. The numerous modern publications about superposition of macroscopic quantum states \cite{Mooij99,Makhlin01,Leggett02,Mooij03,Clark08}, "qubits" entanglement   \cite{Averin98,Makhlin99,Berkley03,Ilichev04,Mooij07,Devoret10} and the violation of the Bell's inequality \cite{Martinis2009,Korotkov2010} imply that their authors {\it give up this program within the realm of the "macroscopic"} in defiance of the firm belief of Einstein that {\it no one is likely to be inclined} to do this. But I think that these authors do not imagine even that they have renounced the real and followed to bishop Berkeley. 

Einstein wrote as far back as 1928 to Schrodinger \cite{Lett1928}: {\it "The soothing philosophy-or religion?-of Heisenberg-Bohr is so cleverly concocted that it offers the believers a soft resting pillow from which they are not easily chased away"}, see the cite on the page 99 of \cite{QuCh2006}. The Bell's remark that the progress of QM {\it is made by sleepwalkers} \cite{Bell1984} witnesses that the {\it soft resting pillow} could lull to sleep some generations of physicists. The Einstein's words turned out prophetic in the sense that most physicists are not conscious that the {\it soothing philosophy-or religion?-of Heisenberg-Bohr} implies the rejection of the "real" as the program of physics. Therefore numerous modern authors are sure that the orthodox QM can describe a process of quantum computation in spite of the reiterated warnings of Heisenberg that it can not describe what happens between measurements. Heisenberg and Bohr could not explaine enough clear the believers the grave aftereffects of the renunciation of the "real" it may be because of their incomplete awareness of the all inevitable consequences. In spite of the logic they did not waive the universality demand of QM. Bohr, for example, defended the universality of the uncertainty and complementarity principles in his discussion with Einstein \cite{Bohr1949}. The orthodox QM developed and was studied as a universal theory of a quantum reality because of a naive  realism inherent to the thinking of most scientists who are not inclined to philosophy. One of the consequences of this misunderstanding are the groundless publications \cite{Mooij99,Makhlin01,Leggett02,Mooij03,Clark08,Averin98,Makhlin99,Berkley03,Ilichev04,Mooij07,Devoret10,Martinis2009,Korotkov2010} and many others. 

Einstein, voicing of the Kantian tradition in the sentence: {\it "The real is not given to us, but put to us (aufgegeben) (by way of a riddle)"}, attempted to explain that the "real" is {\it "a conceptual construction for the grasping of the inter-personal, the authority of which lies purely in its validation"} \cite{Einstein1949}. The neologism {\it be-able as against observ-able} invented by Bell \cite{Bell1975} is an analog of the {\it conceptual construction} by Einstein. The difference in substance between beables and observables is the essence of the intrinsic contradiction of the idea of a quantum computer as a real device. David Mermin, before he {\it learned to stop worrying}, understood that {\it Most theoretical physicists} fail to distinguish between what is measurable and what is existent \cite{Mermin1993}. The false confidence in the reality of quantum computer is a consequence of this lack of understanding that the observation of the EPR correlation can not be interpreted as its existance because of the inadmissible contradiction with local causality that is motivated by relativity theory. The contradiction with local causality is no problem for {\it the newest orthodoxy}. But could the newest orthodoxy or the old orthodoxy provides us an information what quantum system can be used for quantum bit? Bell stated that QM {\it aimed only to describe systematically the response of the apparatus} \cite{Bell1984} according to the old orthodoxy. He spoke ironically: {\it "And what more, after all, is needed for application?"} The irony of it is that the authors of numerous publication abour quantum bit are sure that some observations can guarantee fully that a quantum system is quantum bit in the real. But then the Schrodinger's cat is quantum bit.

\subsection{Could the Schrodinger cat be quantum bit?}
%\label{sec:2.4}
Both the old and newest orthodoxy profess the principle: "We should not raise questions on which we can not answer". This principle could seem reasonable if no one would forget about existence of these questions. The old orthodoxy refused to answer, for example, on the question about cause and time of a radioactive atom decay. Quantum mechanics, following George Gamow, considers this phenomenon using quantum tunneling described with the $\psi $ - function. According to this description the superposition 
$$\Psi_{atom} = \alpha At_{decay}  + \beta At_{no} \eqno{(1)} $$
of decayed $At_{decay}$ and not decayed $At_{no}$ atom {\it "yields the probability that the particle, at some chosen instant, is actually in a chosen part of space (i.e., is actually found there by a measurement of position)"} \cite{Einstein1949}. Einstein accentuate \cite{Einstein1949}: {\it "On the other hand, the $\psi$-function does not imply any assertion concerning the time instant of the disintegration of the radioactive atom"}. He raises the question: {\it "Can this theoretical description be taken as the complete description of the disintegration of a single individual atom?"} and answers {\it "The immediately plausible answer is: No"} \cite{Einstein1949}. Einstein indicates further the positivistic essence of the old orthodoxy: {\it "To this the quantum theorist will reply: … The entire alleged difficulty proceeds from the fact that one postulates something not observable as "real""}. Following Schrodinger he shows that this positivistic point of view implies that a time-instant of the macroscopic event (the mark on a registration-strip, moved by a clockwork) should be considered as no real \cite{Einstein1949}. Schrodinger has shown the same in his paradox with the unfortunate cat \cite{Schrod35D}. The situation, considered by Schrodinger, can be described with $\psi $ - function  
$$\Psi_{cat} = \alpha At_{decay}G_{yes}Fl_{yes}Cat_{dead} \  +$$
$$  \beta At_{no} G_{no}Fl_{no}Cat_{living}   \eqno{(2)} $$
The cat state $Cat_{dead}$, $Cat_{living}$ is entangled with the states of the small flask of hydrocyanic acid $Fl_{yes}$, $Fl_{no}$, the Geiger counter tube $G_{yes}$, $G_{yes}$ and atom $At_{decay}$, $At_{no}$ with the experiment conditions. When one will open the steel chamber and will see the dead cat the $\psi $ - function (2) will collapse to 
$$\Psi_{cat} = At_{decay}G_{yes}Fl_{yes}Cat_{dead} \eqno{(3)} $$
We can draw the conclusion that the cat is dead $Cat_{dead}$ because the hammer has shattered the small flask of hydrocyanic acid $Fl_{yes}$. The hammer has shattered it because the Geiger counter tube has discharged $G_{yes}$. It is has discharged because the atom has decayed $At_{decay}$. Till this each event had a cause. But the atom decay is causless. There is no term to the right of $At_{decay}$ in (3). The cause is absent or we can not know it. The superposition (1) and (2) describes causless phenomenon. The old orthodoxy states that we should not worry about this. But if we will not worry we can come to a false conclusion about possibility of quantum bit. 

In order to understand that it is possible to combine two famous paradoxes. Thereto we will substitute the radioactive atom for the EPR pairs, two spin particles in the singlet state, as well as in the Bohm's version \cite{Bohm1951} of the EPR paradox. We will use also two Stern-Gerlach analysers, two Geiger counter tubes, two flasks of hydrocyanic acid and two cats $CatA$ and $CatB$ (for each experiment). The Geiger counter tubes will be located on the upper trajectory of each particle after its exit from its Stern-Gerlach analyser, so it will discharge when spin up and will not discharge when spin down. The subsequent events will be as well as in the Schrodinger paradox \cite{Schrod35D}. This gedankenexperiment can be described with the $\psi$-function 
$$\Psi_{EPR,cat} = \alpha |\uparrow _{A}>|\downarrow _{B}> CatA_{dead}CatB_{liv} \  + $$
$$\beta |\downarrow _{A}>|\uparrow _{B}> CatA_{liv} CatB_{dead}   \eqno{(4)} $$
with two types of entanglements: because of the conservation law $|\uparrow _{A}>|\downarrow _{B}>$, $|\downarrow _{A}>|\uparrow _{B}>$ and because of the condition of experiment $|\uparrow _{A}> CatA_{dead}$, $|\uparrow _{B}>| CatB_{dead}$, $|\downarrow_{A}> CatA_{liv}$, $|\downarrow _{B}> CatB_{liv}$. The results of observations will be 
$$\Psi_{EPR,cat} = CatA_{dead}CatB_{liv} $$ or 
$$ \Psi_{EPR,cat} = CatA_{liv} CatB_{dead}   \eqno{(5)} $$
when the Stern-Gerlach analysers are oriented in parallel directions. Let imagine that we do not know why the observed states of cats are correlated as well as we do not know the cause of atom decay. Then, the $\psi$-function describing our knowledge would be  
$$\Psi_{EPR,cat} = \alpha CatA_{dead}CatB_{liv}  + \beta CatA_{liv} CatB_{dead}   \eqno{(6)} $$ 
According to the pure positivism, i.e. the old orthodoxy, or the newest orthodoxy the cat's states are entangled. We can even make sure of the violation of the Bell's inequality and have removed all doubts that the cats are quantum bits. 

No one proposed in earnest to use the cats for creation of a quantum computer for the present. Other macroscopic systems are considered as obvious quantum bits in numerous publications \cite{Mooij99,Makhlin01,Leggett02,Mooij03,Clark08}. These publications are direct consequence of the misinterpretation of the QM, created by positivists, as a universal description of a unique reality. The progress made at the description of macroscopic quantum phenomena is {\it immensely impressive} \cite{Bell1984}. But {\it "sleepwalkers"} making this progress have taken no notice of a fundamental diference between this description and the description of atomic phenomena. The theories of macroscopic quantum phenomena were created during the period 1940 -1950s when only the very few paid attention to the controversy between realists and positivists and most physicists ignored the EPR paradox. Therefore this theories are interpreted by the majority as an integral part of the orthodox QM although we should not deny realism at the description of macroscopic quantum phenomena. There is not the fundamental obscurity connected with the $\psi $ - function collapse at observation because the observation in itself can not exert influence on the wave function describing macroscopic quantum phenomena. But there is the fundamental obscurity different in essence from the one revealed by Einstein, Schrodinger, Bell and others. {\it Our theorists stride through that obscurity unimpeded} as well as through the obscurity on which Bell noted \cite{Bell1984}.   

\section{The wave function and the $\psi $ - function}
%\label{sec:3}
The first macroscopic quantum phenomenon, superconductivity was discovered experimentally by Heike Kamerlingh Onnes as far back as 1911, even before the postulation of the Bohr's quantization. The second macroscopic quantum phenomenon, superfluidity was discovered by Pyotr Kapitsa in 1937. Lev Landau has proposed in 1941 a first description of superfluidity phenomena \cite{Landau41}. He proposed also in this paper \cite{Landau41} to describe the superconductivity phenomena with help of a wave function $\Psi = \Psi _{0}\exp(\frac{i}{\hbar }\sum_{a}\chi _{a})$ of the base state of electron liquid, where $\bigtriangledown \chi _{a} = p_{a}$ is the momentum of $"a"$ electron in a point $r$ . This proposal to use a wave function together with the theory of second order phase transitions by Landau \cite{Landau37} are the basis of the famous Ginzburg-Landau theory \cite{GL1950}. According to  the GL theory the order parameter of superconducting state is a wave function $\Psi _{GL} = |\Psi _{GL}|\exp i\varphi $, in which $|\Psi _{GL}|^{2} = n_{s}$ is interpreted as the superconducting electron density and $\hbar \bigtriangledown \varphi = p$ is momentum of single superconducting electron. Landau has written as far back as 1941 the relation for superconducting current density 
$$j_{s} = \frac{n_{s}q}{m}(\bigtriangledown \chi - qA) \eqno{(7)}$$
postulating that all electrons have the same momentum $\bigtriangledown \chi _{a} = \bigtriangledown \chi $. $\bigtriangledown \chi = \hbar \bigtriangledown \varphi = p = mv + qA$ is the canonical momentum of a particle with a mass $m$ and a charge $q$ in the presence of a magnetic vector potential $A$. $(\bigtriangledown \chi - qA)/m = v$ is the velocity. The same relation (7) is obtained in the GL theory for the case when the density is constant in space $\bigtriangledown n_{s} = 0$. The relation (7) can describe quantization effects observed in superconductors. The quantization can be deduced from the requirement that the complex wave function must be single-valued $\Psi _{GL} = |\Psi _{GL}|\exp i\varphi = |\Psi _{GL}|\exp i(\varphi + n2\pi)$ at any point in superconductor. Therefore, its phase must change by integral multiples of $2\pi $ following a complete turn along the path of integration, yielding the Bohr-Sommerfeld quantization 
$$\oint_{l}dl \nabla \varphi = \oint_{l}dl \bigtriangledown \chi /\hbar = \oint_{l}dl p/\hbar  = n2\pi \eqno{(8)}$$ 
According to the relations (7), (8) and  $\oint_{l}dl A = \Phi $ the integral of the current density along any closed path inside superconductor 
$$\mu _{0}\oint_{l}dl \lambda _{L}^{2} j_{s}  + \Phi = n\Phi_{0}  \eqno{(9)}$$  
must be connected with the integral quantum number $n$ and the magnetic flux $\Phi $ inside the closed path $l$. $\lambda _{L} = (m/\mu _{0}q^{2}n_{s})^{0.5} = \lambda _{L}(0)(1 - T/T_{c})^{-1/2}$  is the quantity generally referred to as the London penetration depth; $\lambda _{L}(0) \approx 50 \ nm = 5 \ 10^{-8} \ m$ for most superconductors \cite{Tinkham}; $\Phi _{0} = 2\pi \hbar /q$ is the quantity called flux quantum. 

The postulate used by Landau in order to obtained (7) \cite{Landau41} was absolutely illegal according to the orthodox QM. Electrons are Fermi particles, fermions and can not have the same momentum according to the Pauli exclusion principle. Obviously, this contradiction may be eliminated if electrons can exhibit bosonic behaviour when they become bound in pairs. The valid mechanism of the pairing was proposed in 1957 by Bardeen, Cooper and Schrieffer \cite{BCS1957}. According to the BCS theory \cite{BCS1957} electrons interact through the exchange of phonons, forming Cooper pairs. Numerous experimental results give evidence of the electron pairing. The periodicity in magnetic field of different observables corresponds to the flux quantum $\Phi _{0} = 2\pi \hbar /q$ with the charge $q = 2e$, but no $q = e$ as it followed from the Landau paper \cite{Landau41}.  

According to (9) and the Maxwell equation $curl H = j$ the current density decreases strongly inside superconductor $j_{s} = j_{0}\exp{-\frac{|r - r_{ex}|}{\lambda _{L}}}$ where the coordinate $r$ run from the surface (at $r = r_{ex}$) into the interior $r < r_{ex}$ of the superconductor cylinder with the external radius $r_{ex}$. Pursuant to this screening of the magnetic field the current density (9) or velocity quantization 
$$\oint_{l}dl v  =  \frac{2\pi \hbar }{m}(n - \frac{\Phi}{\Phi_{0}})  \eqno{(10)}$$ 
is observed in a cylinder or ring with narrow wall $w = r_{ex} - r_{in} \ll \lambda _{L}$ when the screening is weak, $r_{in}$ is the internal radius. In the opposite case of the strong screening $w = r_{ex} - r_{in} \gg \lambda _{L}$ a closed path $l$ with a radius $r$, $r - r_{in} \gg \lambda _{L}$ and $r_{ex} - r  \gg \lambda _{L}$, is inside superconductor for which $j_{s} = 0$. Thus, the flux quantization $\Phi = n\Phi _{0}$ should be observed in a superconductor cylinder with the wide wall $w \gg \lambda _{L}$ according to (9). The quantum number $n$ must be equal zero when the wave function $\Psi _{GL} =|\Psi _{GL}|\exp{i\varphi }$ has no singularity inside $l$, since the radius $r$ of the integration path $l = 2\pi r$ can be decreased down to zero in this case. Consequently, the magnetic field must be completely expelled $\Phi = n\Phi _{0} = 0$ from a superconductor without any non-superconducting hole, i.e. at $ r_{in} = 0$. The expulsion of magnetic flux $\Phi $ from the interior of a superconductor, known as the Meissner effect, was discovered by Meissner and Ochsenfeld as far back as 1933 \cite{Meissner1933}. This paradoxical effect is the first evidence of macroscopic quantum phenomena. It may be considered as a particular case $n = 0$ of the flux quantization $\Phi = n\Phi _{0} = 0$. The flux quantization $\Phi = n\Phi _{0}$ discovered experimentally in 1961 \cite{FQ1961} has allowed to measure first the value of the flux quantum $\Phi _{0} = 2\pi \hbar /2e \approx  2.07 \ 10^{-15} \ T m^{2}$. More exactly this value was measured with help of the observation by W. A. Little and R. D. Parks \cite{LP1962} of the resistance oscillations of superconductor cylinder with narrow wall $w \ll \lambda _{L}$. The Little-Parks experiment is first observation of the velocity quantization (10). These and other effects of quantization as well as the Meissner effect are described excellently with help of the GL wave function $\Psi _{GL} = |\Psi _{GL}|\exp i\varphi $. 

The GL theory, in contrast to the BCS theory, is sometimes called phenomenological as it describes some of the phenomena of superconductivity without explaining the underlying microscopic mechanism. This discrimination is based on the delusion that the QM in whole is not phenomenological theory.  The term phenomenology in science is used to describe a body of knowledge which relates empirical observations of phenomena to each other. The orthodox QM  describes just correlation between empirical observations, observables but no beables. It can exceed the limits of the phenomenological level only in some interpretations. The GL theory describes rather beables than observables. In this sense it is less phenomenological than the orthodox QM. But it is important to understand that all quantim theories can only describe but can not explain quantum phenomena. For example, it is mistake to think that the BCS theory can explain the pairing of all electrons considering only very few part $\sim \epsilon _{D}/\epsilon _{F} \approx 0.001$ of theirs.  

\subsection{The Schrodinger's and Born's interpretations of the wave function}
%\label{sec:3.1}
The contradiction of the Schrodinger's wave function with realism has originate from the Born's interpretation. The interpretation proposed by Schrodinger himself was realistic.  Feynman in the Section "The Schrodinger Equation in a Classical Context: A Seminar on Superconductivity" of his Lectures on Physics \cite{FeynmanL} stated that Schrodinger {\it "imagined {\bf incorrectly} that $|\Psi |^{2}$ was the electric charge density of the electron. … It was Born who {\bf correctly} (as far as we know) interpreted the $\Psi $ of the Schrodinger equation in terms of a probability amplitude-that very difficult idea that the square of the amplitude is not the charge density but is only the probability per unit volume of finding an electron there, and that when you do find the electron some place the entire charge is there"}. But further Feynman wrote that {\it in a situation in which $\Psi $ is the wave function for each of an enormous number of particles which are all in the same state, $|\Psi |^{2}$  can be interpreted as the density of particles}. 

This consideration reveals evidently that even Feynman could not take seriously the positivism of Heisenberg, Bohr, Born and other {\it positivistically inclined physicists}. He stated in fact that Schrodinger interpreted his wave function incorrectly and the Born's interpretation is correct in all cases. That is Feynman, as well as most physicists, considered subconsciously QM as an universal theory of a quantum world. He had not observed that the subject of the wave function description changes in essence in his consideration, from  observables to beables. According to the Born's interpretation the Schrodinger wave function describes a probability $|\Psi_{BIn} |^{2}dV$ to observe a particle in a volume $dV$ and must collapse at finding it in a volume. The wave function describing the real density of particles $|\Psi_{ShIn} |^{2} = n$ can not collapse, in accordance with the realistic Schrodinger's interpretation. Thus, the Feynman's statement \cite{FeynmanL} that the Born's interpretation is correct and the Schrodinger's one is incorrect should not be considered universal. 

We must remember that the orthodox QM can describe only phenomena and that such description should not be universal. The Schrodinger's interpretation is obviously inapplicable for the description of some phenomena observed on atomic level. But its interpretation is applicable, more than the Born's one, for the description of other phenomena. For example, the Born's interpretation is quite useless for the description of macroscopic quantum phenomena, in spite of the common belief. The term "wave function" introduced by Schrodinger for the description of {\it a real situation} was carried by {\it the positivistically inclined physicists} to the Born's interpretation. But Schrodinger \cite{Schrod35D}, Einstein \cite{Einstein1949} and other opponents of the positivism used the term "$\psi $ - function" for this case. Thus, according to Schrodinger the wave function describes {\it a real situation} (beables) and can not collapse whereas the $\psi $ - function describes results of observations (observables) and should collapse. I use the terms "wave function" for the case of the Schrodinger's interpretation $\Psi_{ShIn} $  and "$\psi $ - function" for the Born's interpretation $\Psi_{BIn} $ in the same sense.  

\subsection{The Aharonov - Bohm effects observed in the two-slit interference experiment and in mesoscopic ring}
%\label{sec:3.2}
Some effects, seeming similar, differ in essence depending on the presence or the absence of the collapse in its description. One of the most obvious case of this difference is the Aharonov - Bohm effect, i.e. the effect of the electromagnetic potential on the phase of the $\psi $ - function or the wave function. Y. Aharonov and D. Bohm considered this effect for the two-slit interference experiment \cite{AB1959}. The Schrodinger's interpretation is obviously inapplicable for its description and the $\psi $ - function must be used. The interference pattern 
$$P(y) = A_{1}^{2} + A_{2}^{2} + 2A_{1}A_{2} \cos(\Delta \varphi _{1} - \Delta \varphi _{2}) \eqno{(11)} $$ 
observed on a detecting screen at the same velocity $v$ of particles can be described with the superposition $\Psi_{BIn1} = \Psi _{BIn1} + \Psi _{BIn2}$ of two $\psi $ - functions $\Psi _{BIn 1} = A_{1}e^{i\varphi _{1}}$, $\Psi _{BIn2}= A_{2}e^{i\varphi _{2}}$ of momentum $p = mv$ eigenstates. These $\psi $ - functions describe two possible path $l_{1}$, $l_{2}$ through the first slit $\Delta \varphi _{1} = \int_{S}^{y}dr_{1}(p/\hbar ) + tE/\hbar$ and the second slit $\Delta \varphi _{2} = \int_{S}^{y}dr_{2}(p/\hbar ) + t E/\hbar $ between a particle source $S$ and a point $y$ on the detecting screen. $A_{1}$, $A_{2}$ are the amplitudes of the arrival probability at the point $y$ of a particle passing through the first, second slit. Aharonov and Bohm have noted more than fifty years ago \cite{AB1959} that the phase difference $\Delta \varphi _{1} - \Delta \varphi _{2} = \int_{S}^{y}dr_{1}(p/\hbar ) - \int_{S}^{y}dr_{2}(p/\hbar ) = \int_{S}^{y}dr_{1}(mv/\hbar ) - \int_{S}^{y}dr_{1}(mv/\hbar ) + \oint dr (eA/\hbar ) =2\pi  (l_{1} - l_{2})/\lambda _{deB} + 2\pi \Phi /\Phi _{0}$ and consequently, according to (11), the interference pattern should shift with magnetic flux $\Phi $ because of the relation $p = mv + eA$ between canonical momentum $p$ and electron velocity $v$ in the presence of a magnetic vector potential $A$. $\lambda _{deB} = 2\pi \hbar /mv$ is the de Broglie wavelength. 

The periodical dependencies in magnetic flux $\Phi $ with period $\Phi _{0} = 2\pi \hbar /q$ are observed in superconductor \cite{PCScien07,PCJETP07,ABNanoSt2009}, see (9), and normal metal rings \cite{PCScien09} also because of the relation $p = mv + eA$, i.e. the Aharonov - Bohm effect \cite{ABRMP1985}. But this Aharonov - Bohm effect differs in essence from the case of the two-slit interference experiment. The periodicity observed in the mesoscopic rings \cite{PCScien07,PCJETP07,ABNanoSt2009,PCScien09} because of the requirement that the complex wave function must be single-valued $\Psi _{ShIn} = |\Psi _{ShIn}|\exp i\varphi = |\Psi _{ShIn}|\exp i(\varphi + n2\pi)$, see (8). This requirement is violated at the description of the two-slit interference experiment because of the collapse of the $\psi $ - function at the observation of the electron arrival in a point $y$ of the detector screen. The phase difference $\Delta \varphi _{1} - \Delta \varphi _{2} = \int_{S}^{y}dr_{1} \nabla \varphi - \int_{S}^{y}dr_{2} \nabla \varphi = \oint_{l} dr \nabla \varphi $ and the probability (11) can change uninterruptedly with the coordinate $y$ and magnetic flux $\Phi $ thanks to the collapse. This uninterrupted variation provides the interference pattern (11) and its shift with $\Phi $. Thus, although the both Aharonov - Bohm effects result from the $\Phi $ influence on the phase variation along a closed path $\oint_{l} dr \nabla \varphi $, see Fig.1 in \cite{FPP2008}, they differ fundamentally: the first one should be described with the $\psi $ - function whereas the second one should be described with the wave function. 

The ignorance about the fundamental difference between these functions has provoked mistakes. For example, the authors \cite{Strambini} have concluded that electrons can be reflected because of magnetic flux $\Phi $ in the Aharonov-Bohm ring. The full or partial reflection at, for example, $\Phi = 0.5\Phi _{0}$, see Fig.2 in \cite{Strambini}, means that the momentum of all or some electrons changes from $p$ to $-p$ without any real force. Therefore the theoretical result \cite{Strambini} contradicts obviously to the law of momentum conservation \cite{Comment2010}. The contradiction with one of the fundamental laws of physics in such scandalous form is absent in the Aharonov - Bohm effect \cite{AB1959}, although there is a problem with non-local force free momentum transfer \cite{QuCh2006,Nature08,Book1989}, which has provoked controversy \cite{Nature08,Book1989,Boyer00,Boyer02,PRL07}. In spite of the change in the interference pattern no {\it overall} deflection of electrons is observed in the Aharonov-Bohm effect because of magnetic flux \cite{QuCh2006}. The transmission (reflection) probability $P_{tr} = \int dy P(y) = \int dx (A_{1}^{2} + A_{2}^{2}) = 1$ can not depend at all on magnetic flux $\Phi $ contrary to the erroneous theoretical result shown on Fig.2 in \cite{Strambini} and in complete agreement with the conservation law. The obvious mistake made by the authors \cite{Strambini} is consequence of their ignorance about the subject of QM description \cite{WakeUp}. 

\subsection{Physics is empirical science. Why the von Neumann proof is foolish}
%\label{sec:3.3}
Jeffrey Bub quotes in \cite{Bub2010} the Bell's dictum {\it "The proof of von Neumann is not merely false but foolish!"} from \cite{Mermin1993} and states that {\it Bell's analysis misconstrues the nature of von Neumann's claim}. I do not think that Bell misunderstood the essence of the von Neumann's no hidden variables proof. In the dictum quoted in \cite{Bub2010} and \cite{Mermin1993} Bell stated that the assumptions of von Neumann are nonsense {\it When you translate they into terms of {\bf physical disposition}}. There is important to remember that QM has misrepresented basically the {\it physical disposition}. According to the credo of Einstein, scientist (natural) {\it "appears as realist insofar as he seeks to describe a world independent of the acts of perception; as idealist insofar as he looks upon the concepts and theories as the free inventions of the human spirit (not logically derivable from what is empirically given); as positivist insofar as he considers his concepts and theories justified only to the extent to which they furnish a logical representation of relations among sensory experiences} \cite{Einstein1949}. This credo is close to the critical rationalism by Karl Popper, his criterion of demarcation between what is and is not genuinely scientific. 

Heisenberg, Bohr and other {\it positivistically inclined physicists} have refused to be realists. The Bohr's statement {\it "It is wrong to think that the task of physics is to find out how Nature is"} \cite{Bohr1958} implies that physics is no quite natural science. But physics remains empirical science. Physicist must be positivist who {\it considers his concepts and theories justified only to the extent to which they furnish a logical representation of relations among sensory experiences} \cite{Einstein1949}. According to the positivistic belief of Heisenberg, Bohr and others our quantum experiences are manifestation of no objective reality but an interaction between atomic objects and measuring instruments. Bell remarks in the Introduction of \cite{Bell1966} that the {\it additional demands} of von Neumann and others {\it are seen to be quite unreasonable when one remembers with Bohr \cite{Bohr1949} "the impossibility of any sharp distinction between the behaviour of atomic objects and the interaction with the measuring instruments which serve to define the conditions under which the phenomena appear".} Von Neumann, as well as the orthodox QM as a whole, could not describe the process of {\it the interaction with the measuring instruments}. Bell wrote directly about this in \cite{Bell1966} p. 448: {\it "A complete theory would require for example an account of the behaviour of  the hidden variables during the measurement process itself. With or without hidden variables the analysis of the measurement process presents peculiar difficulties"}. Therefore the no-go proof by von Neumann does not concern physics as empirical science. The sharp criticism by Bell of the von Neumann proof should be connected with his criticism of the vagueness of the measurement concept in QM \cite{Bell1990}. Jeffrey Bub \cite{Bub2010} as well as von Neumann do not take into account that an expectation value should depend not only on hidden variables but also on the interaction with the measuring instruments.

Unfortunately, the tendency to forget that physics is empirical science is typical for many theoretical physicists and especially mathematicians. They discuss the puzzles of rather quantum formalism than quantum phenomena. For example, as it is written in \cite{Nikolic2007} {\it "in more advanced and technical textbooks on QM, the wave-particle duality is rarely mentioned. Instead, such serious textbooks talk only about waves, i.e., wave functions $\psi (x, t)$"}.  The wave-particle duality may seem a myth according to some theories or interpretations \cite{Nikolic2007}. But for physics as empirical science it is more important the question: "Could this duality be considered as a myth according to observations?" The wave-particle duality along with the indeterminism are oldest puzzles of quantum phenomena. Einstein, who introduced into the consideration both the wave-particle duality in 1905 \cite{Einstein1905} and indeterminism in 1916 \cite{Einstein1916}, understood in full measure the paradoxicality of these features. Bohr remembers in \cite{Bohr1949} picturesque phrase by Einstein about {\it "ghost waves (Gespensterfelder) guiding the photons"}. The duality can not exist, at least in the reality of the single Universe. But it is undoubtedly observed, for example at the interference of electrons passing one by one through two slits \cite{electro}: each electron is observed on the detecting screen as a particle, but the interference pattern testifies to the wave observation. 

David Deutsch, who is opponent of the positivism, as well as Einstein, Popper and Bell, is sure that the observations of this paradoxical duality give undoubted evidence of multiple universes \cite{DeutschFR}.  Deutsch invented the idea of the quantum computer in the 1970s as a way to experimentally test the "Many Universes Theory" of quantum physics - the idea that when a particle changes, it changes into all possible forms, across multiple universes \cite{Father}. This idea allows to understand why quantum computer may excel the classical one. It can do {\it "a number of computations simultaneously in different universes"} \cite{Father}. The idea of multiple universes seems mad for most physicists. But without multiple universes we should call quantum computer as a real device in question because of the contradiction of the superposition principle and the EPR correlation with the reality of single universe.

In the last years the quantum interference of fullerenes and biomolecules with size up to $3 \ nm$ was observed \cite{ZeilInEx03,ZeilInEx07}. Zeilinger considered \cite{Zeil2004} a possibility of the interference experiment with viruses and nanobacteria. Claus Kiefer interprets these observations as the evidence {\it "that there is no obvious limit to the validity of quantum theory. It may apply to macroscopic systems"} \cite{Kiefer2010}. But it is not correct, at least, that it is possible to observe the quantum interference of macroscopic particles. In order to observe the two-slit interference pattern of particle with size $a$ its period $\lambda _{deB}L/d$ should be larger $a$. A slit width and a distance between slits $d$ can not be smaller the particle size $a$. Therefore the distance $L$ between the double-slit screen and the detector screen should be larger than $L = a^{2}/\lambda _{deB}$. Particles pass this distance during a time $t = L/v$ at a velocity $v$. Therefore the interference experiment should take the time 
$$t_{exp} > \frac{a^{2}}{\lambda _{deB}v} = \frac {g}{2\pi \hbar}a^{5} \approx 1.5 \ 10^{-9} \ \frac{c}{nm^{5}} \times  a^{5} \eqno{(12)}$$ 
since the de Broglie wavelength $\lambda _{deB} = 2\pi \hbar /mv$ and the particle mass $m \approx  ga^{3}$. The value $g/2\pi \hbar \approx 1.5 \ 10^{36} \ c/m^{5} = 1.5 \ 10^{-9} \ c/nm^{5}$ at the typical mass density $g \approx  10^{3} \ kg/m^{3}$ of all matter including viruses and bacteria. Thus, the time of the interference experiment should increase strongly with particle size $a$. The interference observations \cite{ZeilInEx03,ZeilInEx07} of fullerenes and biomolecules with $a \leq 3 \ nm$ did not take a long time $1.5 \ 10^{-9} \ c/nm^{5} \times a^{5} \approx 3.6 \ 10^{-7} \ c$. But the interference observation of viruses with  $a \approx 60 \ nm$ will require appreciably longer time $> 1.5 \ 10^{-9} \ c/nm^{5} \times a^{5} \approx 1 \ c$. The transit of each particle with $a \approx 1840 \ nm = 1.84 \ \mu m = 1.84 \ 10^{-6} \ m$ between the double-slit screen and the detector screen should take the time $\approx 1 \ year \approx 31536000 \ c$; particle with $a \approx 10 \ \mu m$ - the time $\approx 4767 \ year $; with $a \approx 100 \ \mu m = 10^{-4} \ m$ - the time $\approx 4.767 \ 10^{8} \ year $ and so on.

\subsection{Why the macroscopic quantum phenomena can be observed.} 
%\label{sec:3.4}
It is impossible also to observe the Bohr's quantization of a macroscopic particle. Phenomena connected with the spectrum discreteness can be observed when the energy difference   
$$\Delta E_{n+1,n} = \frac{mv_{n+1}^{2}}{2}  - \frac{mv_{n}^{2}}{2} \approx \frac{\hbar ^{2}}{2mr^{2}}(2n+1) \eqno{(13)}$$ 
between permitted states $mv_{n}r = n\hbar $ of a particle with a mass $m$ in a ring with radius $r$ exceeds the energy of thermal fluctuation $\Delta E_{n+1,n} > k_{B}T$. $k_{B} \approx 1.4 \ 10^{-23} \ J/K$ is the Boltzmann constant. The spectrum of atom is strongly discrete because of the small radius $r \approx r_{B} = 0.05 \ nm = 5 \ 10^{-11} \ m$ of atom orbits and the small mass of electron $m_{e} = 9 \ 10^{-31} \ kg$. The energy $\hbar ^{2}/2m_{e}r_{B}^{2} \approx 2 \ 10^{-18} \ J$ corresponding to the very high temperature $T \approx 160000 \ K$ for the radius $r_{B}$ of the first Bohr's orbit falls down to the $ \approx 2 \ 10^{-26} \ J$ corresponding to the very low temperature $T \approx 0.0016 \ K$ at the radius $r \approx 500 \ nm = 5 \ 10^{-7} \ m $ of a nano-ring which can be made now. Therefore the Bohr's quantization of single electron can be observed in such ring only at very low temperature. The persistent current is observed \cite{PCScien09} in normal metal ring with $r \approx 500 \ nm $ at the temperature $T \approx 1 \ K$ because it is created by electrons on the Fermi level of the metal \cite{PC1988} for which the quantum number is great $n \approx n_{F} \approx mv_{F}r/\hbar \approx 10000$, see (13). The results \cite{PCScien09} give evidence of the observation of the Bohr's quantization of electrons even in the ring with the radius much greater than the radius of atom orbits. But we have few chance to observe the discrete spectrum even of virus with $a = 60 \ nm$. For a particle with size $a$ the temperature of the measurement should be 
$$T_{exp} < \frac{\Delta E_{1,0}}{k_{B}} < \frac{\hbar ^{2}}{k_{B}2g}\frac{1}{a^{5}} = 3 \ 10^{-4} \ K \ nm^{5} \ \frac{1}{a^{5}}  \eqno{(14)}$$ 
since the ring radius $r$ must exceed the particle size $a$. The Bohr's quantization of a particle with size $a = 1 \ nm$ can be observed at $T_{exp} < 0.0003 \ K$. The discrete spectrum of virus $a = 60 \ nm$ can be observed only at the much lower temperature $T_{exp} < 3.8 \ 10^{-13} \ K$.

The relations (12) and (14) may be considered as a quantitative expression of the correspondence principle formulated by Bohr as far back as 1920. Macroscopic quantum phenomena can not be observed according to this principle. {\it Our theorists} have stridden through that contradiction unimpeded and have described successfully macroscopic quantum phenomena. It would seem that the description obtained by the theorists does not overstep the limits of the orthodox QM. The explanation of the macroscopic quantum phenomena is connected with the Bose-Einstein condensation. But Landau accentuated in \cite{Landau41} that the Bose-Einstein condensation in itself can not explain the superfluidity phenomenon. In order to describe this phenomenon Landau postulated virtually in \cite{Landau41} that atoms in superfluid $^{4}He$ can not have individual velocity and superfluid condensate moves as one big particle. He obtained in \cite{Landau41} the relation for superconducting current (7) also using this postulate. 

There is important to note that this Landau's postulate should be applied also for superconducting pairs. The energy gap of the BCS theory \cite{BCS1957} concerns the energy spectrum of electrons but not pairs. The GL wave function $\Psi _{GL} = |\Psi _{GL}|\exp i\varphi $ can describe the quantization phenomena (9) observed in superconductor structures \cite{PCScien07,PCJETP07,ABNanoSt2009} thanks to the Landau postulate implied in the interpretation of $|\Psi _{GL}|^{2} = n_{s}$ as the pair density. This interpretation means that all pairs $N_{s} = Vn_{s}$, for example in a ring with the volume $V = s2\pi r$, the radius $r$ and the section $s$, have the same momentum $\hbar \bigtriangledown \varphi = p$. The GL theory could describe macroscopic quantum phenomena thanks to the assumption about two particles: macroscopic one - the condensate $N_{s} = Vn_{s}$ with a big mass $M = N_{s}m$, described with $|\Psi _{GL}|$, and microscopic one - the electron pair $2e$ with momentum $p = mv_{s} + 2eA$, described $p = \hbar \bigtriangledown \varphi $, with the phase $\varphi $ of the GL wave function. The difference between the permitted (10) velocity values $v_{n+1} - v_{n} = \hbar /mr$ depends on the microscopic mass $m$ of single pair. Therefore it is much larger than for virus and any macroscopic particle. In addition, the macroscopic mass $M = N_{s}m$ should be in the relation (13) for the energy difference   
$$\Delta E_{n+1,n} = \frac{Mv_{n+1}^{2}}{2}  - \frac{Mv_{n}^{2}}{2} = \frac{M}{m}\frac{\hbar ^{2}}{2mr^{2}}(2n+1) = $$
$$ = N_{s}\frac{\hbar ^{2}}{2mr^{2}}(2n+1) \eqno{(15)}$$ 
in accordance with the Landau's postulate that all $N_{s} = Vn_{s}$ pairs have the same quantum number $n$ and none of the $N_{s}$ pairs can change $n$ individually. The energy difference (15) increases with the increase of superconductor sizes $\propto N_{s} = n_{s}s2\pi r$. The discreteness becomes stronger in macroscopic superconductor even in spite of the decrease of velocity discreteness $v_{n+1} - v_{n} = \hbar /mr$ with the radius $r$ increase: $\Delta E_{n+1,n} \approx  n_{s}s2\pi r(\hbar ^{2}/2mr^{2}) \propto  (s/r) = hw/r$, where $h$ and $w$ are the height and width of a ring or cylinder, for example.

Thus, in order to describe macroscopic quantum phenomena a new duality should be postulated. This duality of microscopic - macroscopic particles differs in essence from the wave-particle duality introduced by Einstein \cite{Einstein1905} and especially from the duality of the Born's interpretation. The GL wave function, as direct opposed to the $\psi $ - function, can not collapse at observation. There is not any contradiction with realism at the description of macroscopic quantum phenomena. One may say that the GL wave function is real, to be more precise it describes reality. Feynman wrote in \cite{FeynmanL} that at the description of superconductivity {\it the wave functions take on a physical meaning which extends into classical, macroscopic situations}. Unfortunately even brilliant physicist interpreted his renunciation of realism only as the deliverance from classical prejudices. Feynman was among the theorists who strode unimpeded through the fundamental obscurity in QM on which Bell indicated \cite{Bell1984}. He was sure that Einstein - Podolsky - Rosen paradox is not paradox at all \cite{FeynmanL}. It may be therefore he could not realised the fundamental difference between the wave function describing superconductivity and the $\psi $ - function. 

The EPR paradox \cite{EPR1935} has revealed that the $\psi $ - function can not be applied for description of local reality because of its collapse at observation. The GL wave function describing the real density $|\Psi _{GL}|^{2} = n_{s}$ can not change at all because of any observation in itself.  The description of macroscopic quantum phenomena does not contradict realism, at least in the sense of the EPR paradox. There is not the fundamental obscurity indicated by Bell and the problem discussed now by experts \cite{Norsen09,Ghirardi10}. The GL wave function can not describe superposition of states, which should collapse at observation, and, consequently, the EPR correlation. This absence of problems with local realism means, in particular, that superconductor structures should not differ from other macroscopic object, for example cats, with respect to their utilization for the creation of quantum computer. This logical deduction would seem paradoxical for the authors of the numerous publications about superconducting quantum bits \cite{Mooij99,Makhlin01,Leggett02,Mooij03,Clark08,Averin98,Makhlin99,Berkley03,Ilichev04,Mooij07,Devoret10}. I should note that these publications result from the misinterpretation by their authors of the superposition essence.

\subsection{Must we use the superposition principle for description of macroscopic quantum phenomena?}
%\label{sec:3.5}
This misinterpretation has resulted from the dual attitude to this principle. Most physicists learned during some ten years that it is the cardinal positive principle of the QM \cite{LandauL}. Because of this history many authors \cite{Leggett02,Mooij03,Clark08} are sure that an experimental result is the evidence of superposition if it {\bf can} be described with help of the superposition principle. They ignore totally the irremediable contradiction of this principle with realism. The cardinal positive principle of the QM can describe results of observations, according to the old orthodoxy, or our knowledge, according to the newest orthodoxy. But it can not be interpreted as a description of {\it any real situation}. It is impossible both because of the logic: nothing can be in different places simultaneously and no parameter can has different values at the same time and because of the irremediable conflict with local realism revealed with the EPR correlation \cite{EPR1935}. The only feasible realistic interpretation of superposition must presuppose many Universes. According to Deutsch, {\it "quantum superposition is, in Many Universes terms, when an object is doing different things in different universes"} \cite{Father}. 

The Bell's {\it attitude to the Everett-de Witt 'many world' interpretation} was {\it a rather negative} \cite{Bell1987,Bell1976}. His hidden-variables model for a single spin - 1/2 \cite{Bell1966} has demonstrated that even the very paradoxical Stern-Gerlach effect \cite{SternGerlach} can be describe in our habitual reality of single Universe. The paradoxical nature of this effect was realised \cite{Einstein1922} just after its discovery in 1922. Bohr wrote about that time \cite{Bohr1949}: {\it "In the following years, during which the atomic problems attracted the attention of rapidly increasing circles of physicists, the apparent contradictions inherent in quantum theory were felt ever more acutely. Illustrative of this situation is the discussion raised by the discovery of the Stern-Gerlach effect in 1922. On the one hand, this effect gave striking support to the idea of stationary states and in particular to the quantum theory of the Zeeman effect developed by Sommerfeld; on the other hand, as exposed so clearly by Einstein and Ehrenfest \cite{Einstein1922}, it presented with {\bf unsurmountable difficulties} any attempt at forming a picture of the behaviour of atoms in a magnetic field"}. Indeed, it seems impossible to describe realistically the experimental results \cite{SternGerlach} according to which magnetic moment has an identical value of projection on any direction. Nevertheless Bell has formed a picture using the quantum postulate proposed by Bohr \cite{Bohr1928} in order to show that the no-go theorem of von Neumann is unreasonable. 

The no-go theorems may be interpreted as the argument against any possibility to describe a quantum phenomena without the superposition principle, i.e. realistically: we must use this principle for description of a quantum phenomena if it {\bf can not} be described in a different way. The only attempt to propose a no-go theorem for macroscopic quantum phenomena is known \cite{Clark08,LGI2010} as the Leggett-Garg inequality \cite{Leggett1985}. The inequality (2) in \cite{Leggett1985} seems formally similar to the Bell's inequality \cite{Bell1964}. But, as L. E. Ballentine noted as far back as 1987 \cite{Ball1987}, the analogy between the Bell-type inequalities and the Leggett-Garg inequalities misleads. The locality postulate, playing a key role in the Bell's no-go theorem \cite{Bell1964}, can not be applicable to the single localised system considered in \cite{Leggett1985}. Leggett and Garg state in the Comment \cite{Leggett1987} that the assumption (A2) in \cite{Leggett1985} (Noninvasive measurability at the macroscopic level) {\it plays the same role as that  of locality in Bell's theorem}. But the impossibility to measure a variable without its inevitable change at the measurement can not be the basis of any no-hidden-variables theorem. Variables are hidden just because of the impossibility of the noninvasive measurability. The no-hidden-variables theorem by Leggett and Garg \cite{Leggett1985} is unreasonable because of the same reason as that of the one by von Neumann: the orthodox QM predicts results of {\it the interaction with the measuring instruments} but it can not describe the process of this interaction.

Moreover there is no necessity at all to use the superposition principle for a description of any phenomena which can observed at measurements of the rf SQUID (i.e. single superconducting loop interrupted by Josephson junction) considered in \cite{Leggett1985}. Superconducting state of the loop can be describe with help of the GL wave function which can not collapse at any observation. Therefore the description of the rf SQUID in itself can not contradict macroscopic realism. The doubt about reality of the magnetic flux is provoked in the paper \cite{Leggett1985} because of  the utilization of the $\psi $ - function, in addition to the wave function. The $\psi $ - function can apply speculatively for description of state superposition of any macroscopic object, for example the cat \cite{Schrod35D} or the moon \cite{Mermin1993}. But we must not call realism in question without irrefutable empirical evidence, which was not obtained at experimental investigation of the superconducting loop. Moreover there is a valid doubt that such evidence can be obtained some time. The $\psi $ - function used in \cite{Leggett1985} describes the superposition of two states with equal and opposite directed persistent current $I_{p}$. Such a two-states system is represented often as a "particle" of spin 1/2 \cite{Leggett02}. But the loop, in contrast to atom or electron, is not a central-symmetrical three-dimensional system, having three projections $x$, $y$, $z$. The persistent current $I_{p}$ circulating in the $x-y$ loop plane (see Fig.2a in \cite{Clark08}) can induce a magnetic moment $M_{m} = SI_{p}$ and an angular momentum $M_{p} = (2m_{e}/e)M_{m}$ only in the z-direction. The paradoxic feature of the Stern-Gerlach effect can not observed obviously in this one-dimensional system. Consequently there is not even a pretext to use the superposition principle. All experimental results can be described even without hidden variables, in contrast to the Stern-Gerlach effect. 

In spite of the absence of any valid doubt in the reality of the additional flux $\Delta \Phi _{Ip} = LI_{p}$ induced with the persistent current $I_{p}$ the authors of numerous publications \cite{Mooij99,Makhlin01,Leggett02,Mooij03,Clark08} are sure that the superconducting loop interrupted by one or three Josephson junctions can be used as quantum bit, flux qubit. In additional to the groundlessness of this confidence it contradicts evidently to the fundamental law of angular momentum conservation \cite{Nat2009,QI2009}. The $\psi $ - function \cite{Clark08} represents the superposition of two permitted states with the macroscopic persistent current $I_{p} \approx 0.5 \ \mu A = 5 \ 10^{-7} \ A $ circulating clockwise in the one state and anti-clockwise in other one along the macroscopic loop with the area $S \approx 1 \ \mu m^{2} = 10^{-12} \ m^{2}$  \cite{Mooij03}. The energy $\propto I_{p}^{2}$ of these states is equal but the magnetic moment $M_{m} = SI_{p}$ and the angular momentum $M_{p} = (2m_{e}/e) SI_{p}$ differ on the macroscopic values $\Delta M_{m} \approx 10^{5} \ \mu_{B}$, $\Delta M_{p} \approx 10^{5} \ \hbar$ \cite{Nat2009}, where $\mu_{B}$ is the Bohr magneton and $\hbar$ is the reduced Planck constant. The interpretation \cite{Clark08} of the energy-level splitting \cite{Mooij03}, the Rabi oscillations, Ramsey interference \cite{Tanaka06,Semba06,Tanaka07} and other experimental results as evidence of the superposition of the states with macroscopically different angular momentum contradicts inadmissibly to the conservation law. The causeless change of angular momentum can be observed in some quantum phenomena, see the example in \cite{QI2009}. But this change must be in the limits of the uncertainty relation and can not exceed the Planck constant $\hbar$. Thus, the numerous publications about flux qubit conflict with the universally recognized quantum formalism. This conflict is a consequence of the formal utilization of this formalism, without any serious reflection about the subject of its description.

\section{The fundamental obscurity which was not discussed for the present}
%\label{sec:4}
The fundamental obscurity discussed by experts \cite{Nikolic2007,Heisenberg1925,Einstein1949,Heisenberg1958,Bohr1958,Bohr1949,Hooft2007,Einstein1927,EPR1935,Schrod35D,Schrod35E,Entangl,Norsen09,Ghirardi10,Cramer1986,Bell1964,Ballantine87,QuCh2006,Mermin1985,Bell1987,Bell1990,Mermin2006,Ghirardi2008,Mermin2008,Bell1984,Zeil1999,Kiefer2010,Bohr1928,Heisenberg1927,Popper,Bohr1935,Neumann1932,Mermin2001,Mermin2003} is connected with the utilization of the $\psi $ - function at the description of some microscopic quantum phenomena. This obscurity because of a fuzzy status of measurements \cite{Bell1987,Bell1990} is absent at the description of macroscopic quantum phenomena. The measurement process in itself can not change the wave function describing the real density $|\Psi _{GL}|^{2} = n_{s}$. Therefore the problems, revealed with  the EPR paradox \cite{EPR1935}, the Schrodinger's cat paradox \cite{Schrod35D}, the Bell's no-go theorem \cite{Bell1964} and so on, are absent at the macroscopic level. Macroscopic quantum phenomena and their description generate other fundamental puzzles. One may say that these puzzles are more real. In order to be blind to the fundamental obscurity in QM both the old and newest orthodoxy use the information interpretation. This interpretation is natural for the positivistic Born's interpretation of the Schrodinger wave function {\it in terms of a probability amplitude} \cite{FeynmanL}. If {\it the square of the amplitude is not the charge density but is only the probability per unit volume of finding an electron there} \cite{FeynmanL} then it is naturally enough to connect the amplitude $\Psi_{BIn}$ with {\it our knowledge} which is different before and after the {\it finding an electron there}. But the information interpretation should not be considered universal, at least. It can not be valid at the description of many quantum phenomena, first of all the macroscopic one. These phenomena and their description testify against the firm belief of {\it the quantum computation crowd, for many of whom quantum mechanics is self-evidently and unproblematically all about information} \cite{Mermin2001}. The consideration of these phenomena prohibits {\it to stop worrying and love Bohr} \cite{Mermin2003}.

\begin{figure}[b]
\includegraphics{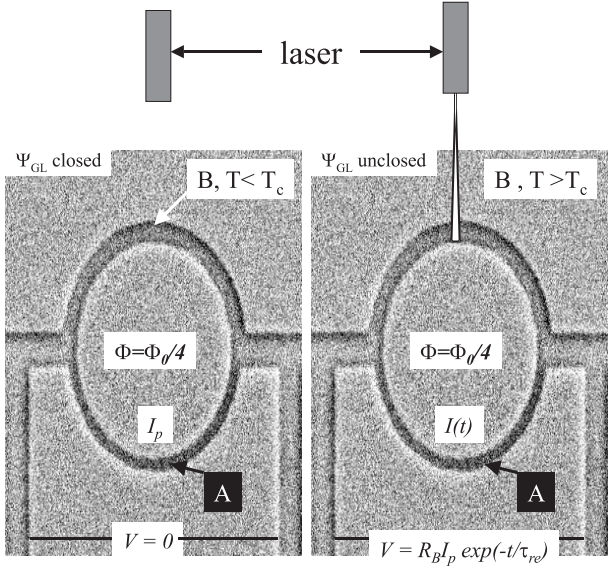}
\caption{\label{fig:epsart} Superconducting loop can be switched between the states with different connectivity of the wave function $\Psi _{GL} = |\Psi _{GL}|\exp i\varphi $ with a real physical influence, for example turning on (the right picture) and turning off (the left picture) of the laser beam heating the loop segment $B$ above $T_{c}$. The persistent current, equal $I_{p} = -s2en_{s}(2\pi \hbar/lm4)$ in a symmetric loop $l$ with the same section $s$ and pair density $n_{s}$ along the whole  $l$, flows at the magnetic flux $\Phi = \Phi_{0}/4$ inside $l$ when the wave function is closed (the left picture). The current circulating in the loop should decay during the relaxation time $\tau_{RL} = L/R_{B}$ after the transition in the state with unclosed wave function because of a non-zero resistance $R_{B} > 0$ of the $B$ segment in the normal state (the right picture). The photo of a real aluminum loop is used in order to exhibit that the gedankenexperiment can be made real.}
\end{figure}

\subsection{The fundamental obscurity connected with phenomena described with the quantum formalism}
%\label{sec:4.1}
Most obvious puzzle may be connected just with the real density $|\Psi _{GL}|^{2} = n_{s}$ described with the wave function $\Psi _{GL} = |\Psi _{GL}|\exp i\varphi $. It can not change because of our look. But we can alter the density of the Cooper pairs with help of a real physical influence. For example, we can decrease this density $n_{s} = n_{s,0}(1 - T/T_{c})$ heating a loop segment. Before the heating the persistent current $I_{p} = s2en_{s}v$ must flow along the superconducting loop with narrow wall $w \ll \lambda $ at $T < T_{c}$ and, for example, at the magnetic flux inside the loop $\Phi = \Phi_{0}/4$, Fig.1, because of the prohibition (10) of the zero velocity $v = 0$ at $\Phi \neq n\Phi_{0}$. The velocity in the permitted state $n = 0$ with minimum energy $\propto (n - \Phi /\Phi_{0})^{2}$ at $\Phi = \Phi_{0}/4$ should equal $v = -2\pi \hbar/lm4$ in a symmetric loop $l$ with the same section $s$ and pair density $n_{s}$ along the whole  $l$, according to (10) and the demand of the current $I_{p} = s2en_{s}v$ continuity. The prohibition will disappear when any segment $B$ is overheated at $t = t_{on}$ above the temperature $T_{c}$ of superconducting transition $T > T_{c}$, for example with help of the laser beam, Fig.1. The current should decay $I(t) = I_{p}\exp -(t- t_{on})/\tau_{RL}$ during the relaxation time $\tau_{RL} = L/R_{B}$ because of a non-zero resistance $R_{B} > 0$ of the $B$ segment in the normal state, see the right picture on Fig.1. The loop should return to the initial state with the closed wave function after turning off of the laser beam at a time $t_{off}$ and the cooling of the $B$ segment down to the initial temperature $T < T_{c}$, see the left picture on Fig.1.  

Thus, one can switch superconducting loop between the states with different connectivity of the wave function $\Psi _{GL} = |\Psi _{GL}|\exp i\varphi $. Such real switching is not possible for $\psi $ - function which can not be interpreted as a description of a {\it real situation}. The closed and unclosed wave functions $\Psi _{GL}$ describe the real situations which differ really one from another: the real persistent current and the velocity $v = -2\pi \hbar/lm4 \neq 0$ of the Cooper pairs is observed in the $\Psi _{GL}$ closed state, whereas this current is absent and $v = 0$ in the $\Psi _{GL}$ unclosed state after the relaxation $t - t_{on} \gg \tau_{RL}$. The velocity change from the quantified value $v = -2\pi \hbar/lm4 \neq 0$ to $v = 0$ occurs in accordance with the Newton's second law $mdv/dt = 2eE$, under the influence of the real force $F_{E} = 2eE$ of electric field $E = - \bigtriangledown V$ acting on each pair $2e$. The potential electric field $E = - \bigtriangledown V \approx V_{B}/(l-l_{B})$ appears because of the potential difference 
$$V_{B} = R_{B}I(t)= R_{B}I_{p}\exp -\frac{t - t_{on}}{\tau_{RL}} \eqno{(16)} $$ 
on the $B$ segment. But there is no real force which could be associated with the change of the velocity from $v = 0$ to $v = -2\pi \hbar/lm4 \neq 0$ and of the angular momentum $\oint_{l}dl p = \oint_{l}dl (mv + 2eA) = m\oint_{l}dl v + 2e\Phi $ of each Cooper pair $2e$ from $2e\Phi $ to $n2\pi \hbar $. Nevertheless such force-free angular momentum transfer 
$$n2\pi \hbar - 2e\Phi = 2\pi \hbar (n - \frac{\Phi }{\Phi_{0}})  \eqno{(17)}$$ 
should occur according to the universally recognized quantum formalism corroborated with numerous experimental results. 

The observations of the quantum periodicity in different parameters, connected with the quantum periodicity in the persistent current $I_{p}(\Phi /\Phi_{0})$, such as the resistance $\Delta R \propto  I_{p}^{2}$ \cite{LP1962,PCJETP07,ABNanoSt2009,Letter07,toKulik2010}, the magnetic susceptibility $\Delta \Phi _{Ip} = LI_{p}$ \cite{PCScien07}, the critical current $I_{c}(\Phi /\Phi_{0}) \propto I_{c0} - 2|I_{p}(\Phi /\Phi_{0})| $ \cite{JETP07J} and others give unequivocal evidence of the velocity quantization (10) and the strong discreteness $\Delta E_{n+1,n} \gg k_{B}T$ of the permitted state spectrum of any real superconductor loop described with the relation (15). The quantum formalism \cite{Tinkham} describes this quantum periodicity as a consequence of the change with the $\Phi /\Phi_{0}$ value of the permitted state $n$ giving the contribution $\propto  \exp -E_{n}/k_{B}T$ of overwhelming size in the measured parameters. But it is not correct to think that the quantum formalism can explain these phenomena. The resistance $\Delta R(\Phi /\Phi_{0}) $ oscillations \cite{LP1962,PCJETP07,ABNanoSt2009,Letter07,toKulik2010} and the magnetic susceptibility $\Delta \Phi _{Ip}(\Phi /\Phi_{0})$ oscillations \cite{PCScien07} at $T \approx T_{c}$ where the loop resistance $R_{l} > 0$ reveal the experimental puzzle. An electric current in a resistive $R_{l} > 0$ loop $l$ should rapidly decay $I(t) = I_{0}\exp (-t/ \tau_{RL})$ in the absence of the Faraday's electric field $E = -dA/dt = -l^{-1} d\Phi /dt$. The current in the aluminum ring with radius $r \approx 1 \ \mu m$ and the inductance $L \approx 10^{-11} \ H$ used in \cite{PCScien07,Letter07,toKulik2010} should decay very quickly  $\tau_{RL} = L/R_{l} \approx 10^{-9} \ s$ even at a small resistance $R_{l} \approx 0.01 \ \Omega $. But the observations of the additional magnetic flux $\Delta \Phi _{Ip} = LI_{p}$ \cite{PCScien07} and the additional resistance $\Delta R \propto  I_{p}^{2}$ \cite{Letter07,toKulik2010} at $T \approx T_{c}$ reveal that the persistent current can not decay for a long time even at $R_{l} \gg 0.01 \ \Omega $. 

The quantum formalism can describe even this puzzle if one takes into account the change of the pair angular momentum (17) at the closing of the wave function $\Psi _{GL} = |\Psi _{GL}|\exp i\varphi $ in the loop. It is natural to describe this puzzle as a consequence of the switching between superconducting states with different connectivity of the wave function \cite{PRB2001} because the persistent current $I_{p} \neq 0 $ of Cooper pairs is observed at non-zero resistance $R_{l} > 0$ only in a narrow temperature region near superconducting transition $T \approx T_{c}$, where thermal fluctuations can switch loop segments between superconducting $n_{s} > 0$, $R = 0$ and normal $n_{s} = 0$, $R > 0$ states. Below this region, at $T < T_{c} - \Delta T_{c}/2$, the pair density is non-zero  $n_{s} > 0$ continually and therefore $I_{p} \neq 0 $ but  $R_{l} = 0$. Above this region, at $T > T_{c} + \Delta T_{c}/2$, $n_{s} = 0$, $I_{p} = 0 $, $R_{l} \approx R_{ln}$  continually. $R_{ln} = \rho_{n}l/s $ is the loop resistance in the normal state; $\Delta T_{c}$ is the width of the resistive transition, i.e. the temperature region where $\delta R < R_{l} < R_{ln} - \delta R $, which may depend on the accuracy $\delta R $ of resistance measurement. 

The resistive transition $0 < R_{l} < R_{ln} $ of a homogeneous loop is observed because of small energy difference between normal and superconducting states at $T \approx T_{c}$ \cite{Tinkham}. Thermal fluctuations \cite{Tinkham} switch loop segments between superconducting and normal states at $T \approx T_{c}$ instead of the laser beam with a high probability $\propto \exp -\Delta F_{GL}/k_{B}T$ because of small difference of the GL free energy $\Delta F_{GL} \leq  k_{B}T$ of superconducting $n_{s} > 0$ and normal $n_{s} = 0$ states. The persistent current is observed at $R_{l} > 0$ \cite{PCScien07,Letter07,toKulik2010} because of the wave function closing from time to time with a frequency $\omega _{sw} = N_{sw}/\Theta  $. The angular momentum of each Cooper pair should return to the quantified value $n2\pi \hbar $ at each of the $N_{sw}$ closings during a long time $ \Theta $, changing every time on $2\pi \hbar (n - \Phi /\Phi_{0})$ (17) or less value. This change of the momentum of Cooper pair because of the quantization at the $\Psi _{GL}$ closing in a time unit 
$$\oint_{l}dlF_{q}=2\pi \hbar (\overline{n}- \frac{\Phi }{\Phi_{0}})\omega _{sw}  \eqno{(18)}$$
was called in \cite{PRB2001} "quantum force". The quantum force (18) as well as the persistent current \cite{PCScien07,Letter07,toKulik2010} is not zero at $\Phi \neq n\Phi_{0}$ and $\Phi \neq (n + 0.5)\Phi_{0}$ and changes periodically with magnetic flux $F_{q}(\Phi /\Phi_{0})$, $I_{p}(\Phi /\Phi_{0})$ because of the strong discreteness $\Delta E_{n+1,n} \gg k_{B}T$ (15) even in the fluctuation region at $T \approx T_{c}$. The observations of the periodicity \cite{PCScien07,Letter07,toKulik2010} give unequivocal evidence that the average value of the quantum number $\overline{n} =  \Sigma _{n} P_{n}n $ equals approximately the integer number $n$ corresponding to lowest energy $\propto (n - \Phi /\Phi_{0})^{2}$. The quantum force $\oint_{l}dlF_{q}$ takes the place of the Faraday's voltage $-d\Phi /dt$ which maintains $IR_{l} = -d\Phi /dt$ a conventional current $I$ circulating in a loop and can describe why the persistent current can not decay $\overline{I_{p}}R_{l} = \oint_{l}dlF_{q}/2e$ in spite of the power dissipation $\overline{I_{p}^{2}R_{l}}$.

The utilization of the quantum force in \cite{PRB2001} for description of the Little-Parks effect \cite{LP1962}, i.e. the first observation of  $I_{p} \neq 0 $ at $R_{l} > 0$, does not overstep the limits of the universally recognized quantum formalism, according to which the change of the angular momentum (17) should take place. The quantum force as well as the quantum formalism can not explain why the change (17) can take place. It is the puzzle which appeared as far back as 1933 when Meissner and Ochsenfeld \cite{Meissner1933} observed first that a superconductor, placed in a weak magnetic field, completely expels the field from the superconducting material except for a thin layer $\lambda _{L} \approx 50 \ nm = 5 \ 10^{-8} \ m$ at the surface. The quantum formalism describes the Meissner effect as the particular case $n = 0$ of the quantization (9), see the Section 3. But it did not make even an attempt to explain why the angular momentum can change at the Meissner effect. At the $\Psi _{GL}$ closing, considered above, this paradoxical change  (17) can not exceed $2\pi \hbar (n - \Phi /\Phi_{0}) < 2\pi \hbar 0.5$  irrespective of the loop $l = 2\pi r$ radius $r$ and the $\Phi = B\pi r^{2}$ value. The Meissner effect can be observed at any radius $r$ of superconductor and $B < B_{c1}$. The angular momentum of each pair can change on a macroscopic value $2\pi \hbar (- \Phi /\Phi_{0}) = 2\pi \hbar (-B \pi r^{2}/\Phi_{0}) \approx  \hbar \ 10^{15}$ at the first critical field $B_{c1} \approx  0.1 \ T$ and the superconductor radius $r = 1 \ m$. This obscurity is macroscopic in truth because of the angular momentum change of all $N_{s} = n_{s}\pi r^{2}h > 10^{29}$ pairs in the cylindrical superconductor. {\it Our theorists stride through that obscurity unimpeded... sleepwalking?} \cite{Bell1984}. Jorge Hirsch wonders fairly that {\it "the question of what is the 'force' propelling the mobile charge carriers and the ions in the superconductor to move in direction opposite to the electromagnetic force in the Meissner effect was essentially never raised nor answered"} \cite{Hirsch2010}. 

Hirsch proposes an explanation of the Meissner effect puzzle \cite{Hirsch2010}. But some consequences of this explanation, for example the electric field inside the superconductor, the relation (23) in \cite{Hirsch2010}, seem unacceptable. In addition, this explanation can not be applied to the angular momentum change at the $\Psi _{GL}$ closing. Therefore we ought conclude that this puzzle can describe but can not be explain, as well as many quantum phenomena. According to the point of view by Hirsch \cite{Hirsch2003,Hirsch2007} the force-free momentum transfer indicates a fundamental problem with the conventional theory of superconductivity \cite{BCS1957}. I think that it indicates a fundamental problem rather with QM as a whole than with a theory of superconductivity. The persistent current $I_{p} \neq 0$ is observed at $R > 0$ not only in superconductor \cite{PCScien07,Letter07,toKulik2010} but also in normal metal rings \cite{PCScien09}. In order to dodge the obvious puzzle the authors \cite{PCScien09} and the author \cite{Birge2009} claim that the electric current can flow in realistic normal metal rings containing atomic defects, grain boundaries, and other kinds of static disorder without dissipating energy. They do not try even to explain how a dissipationless current of electrons can be possible at electron mean free path shorter than the circle length of their rings \cite{PCScien09}. The authors \cite{PCScien09} find a pretext for the dropping of the obvious puzzle using {\it a familiar analog in atomic physics: a current circulating around the atom} although the exponential decrease of the persistent current amplitude with temperature increase, which they observed (Fig.3 in \cite{PCScien09}), testifies against this analog and to fundamental differences between application of some quantum principles on atomic and mesoscopic levels \cite{FFP8}. Igor Kulik, who has described the possibility of  $I_{p} \neq 0$ at $R > 0$ as far back as 1970 both in superconductor \cite{Kulik1970s} and normal metal rings \cite{Kulik1970n} made forty years ago reasonable statement that the taking into account of a dissipation should not result in the disappearance of the persistent current. 

The absence of the universality in application of some quantum principles may be connected with fundamental difference between the wave function and the $\psi $ - function and also with difference of our experimental opportunities on atomic and macroscopic (mesoscopic) levels. The puzzle of the force-free momentum transfer in the Aharonov - Bohm effect observed in the mesoscopic rings \cite{PCScien07,Letter07,toKulik2010,PCScien09} is more real than at the description of this effect observed in the two-slit interference experiment. A partisan of the old or newest orthodoxy may say: "We should not worry about the non-local force-free momentum transfer in the description of the two-slit interference experiment because the $\psi $ - function describes only our knowledge. There can not be a problem with the conservation law in our knowledge". But this trick can not be valid in the case of the persistent current described with the wave function. We can switch the wave function between states with different connectivity, Fig.1, as opposed to the $\psi $ function. This real action should lead to real results which can not be describe completely. The quantum formalism (10) predicts that the velocity of Cooper pairs should change in the point $A$ after closing the wave function in the point $B$, Fig.1. But no one can say how quickly the situation in $A$ will change after the change of the situation in the spatially separated point $B$, Fig.1. Quantum formalism passes over the matter of the phenomenon cause in silence, as usual. An answer on this puzzle can be obtained experimentally. It is more easy to verify experimentally the induction of the potential difference $V_{B} = R_{B}I(t)$ (16) with direct component 
$$V_{dc} = \frac{1}{\Theta }\int_{\Theta }dtV_{B}(t) \approx  L\omega _{sw} \overline{I_{p}}; \  at \ \omega _{sw}\tau_{RL} \ll 1  \eqno{(19a)}$$
$$  V_{dc} \approx  R_{B} \overline{I_{p}}; \ at \ \omega _{sw}\tau_{RL} \gg 1 \eqno{(19b)}$$
with help of repeated switching with a frequency $\omega _{sw} $ of the $B$ segment, Fig.1, between superconducting and normal states \cite{JLTP1998}. The sign and value of the dc voltage (19) should vary periodically with magnetic flux $V_{dc}(\Phi /\Phi_{0})$ like the persistent current $I_{p}(\Phi /\Phi_{0})$ \cite{PCScien07}. Such quantum oscillations of the dc voltage $V_{dc}(\Phi /\Phi_{0}) \propto I_{p}(\Phi /\Phi_{0})$ were observed on segments of asymmetric aluminium rings when the switching take place because of a noise \cite{Letter07,toKulik2010,PerMob2001} or the ac current \cite{PCJETP07,Letter2003}. 

The experimental results \cite{Letter07,toKulik2010} give unequivocal evidence that the persistent current can flow against the dc electric field $E = -\bigtriangledown V$. This puzzle in the observation makes meaningless \cite{toKulik2010} the preposterous claim by the authors \cite{PCScien09,Birge2009} that the persistent current can flow through resistors without dissipating energy. The authors \cite{PCScien09,Birge2009} turn a blind eye also to the other puzzle, which may be connected with the previous one. The author \cite{Birge2009} notes  {\it "time-reversal symmetry should forbid a current choosing one direction over the other around the ring"}. The authors both \cite{PCScien09} and \cite{Birge2009} are sure that the breach of the time-reversal symmetry with a magnetic field can allow the current to choose one direction. But it could allow to choose a direction depending only on the magnetic field direction whereas the direction of the persistent current $I_{p}(\Phi /\Phi_{0})$ \cite{PCScien07,PCScien09} and the dc electric field $E(\Phi /\Phi_{0}) = -\bigtriangledown V_{p}(\Phi /\Phi_{0})$ \cite{Letter07,toKulik2010} changes also with the magnetic field value at $\Phi = n\Phi_{0}$ and $\Phi = (n+0.5)\Phi_{0}$. Each physicist must understand that the observation of the direction change with the value change is a puzzle which may have a fundamental importance \cite{FQMT2004,QTRF2007}. Such puzzle was not observed on the atomic level. In order to observed the Aharonov - Bohm effect for the first Bohr orbit $r_{B} \approx 5.3 \ 10^{-11} \ m$ very high magnetic fields $\Phi_{0}/ \pi r_{B}^{2} \approx 5 \ 10^{9} \ G$, inaccessible for the present, is needed. In additional, we should remember about the fundamental difference between the wave function, describing the persistent current phenomena, and the $\psi $ - function, describing the atomic orbits. The {\it familiar analog} of the persistent current with {\it a current circulating around the atom} used by the authors \cite{PCScien09} indicates once again that most physicists can not take seriously the positivism of Heisenberg, Bohr and others. Although Heisenberg and other founding fathers understood as far back as 1925 \cite{Heisenberg1925}, that any realistic interpretation of electronic ground states of atom is provoked serious objections, the authors \cite{PCScien09} suppose that {\it nonzero orbital angular momentum} can be connected with {\it a current circulating around the atom}. They are apparently ignorant of the Bell's no-hidden-variables theorem according to which any realistic interpretation of orbital angular momentum of atom contradicts to the orthodox QM.

\subsection{Experimental results which can not be describe with help of the quantum formalism}
%\label{sec:4.2}
The puzzles considered above testify to the Bell's remark that the quantum formalism was created by {\it "sleepwalkers"} \cite{Bell1984}. This formalism has been enormously successful in describing various quantum phenomena. Most physicists believe that it can describe all experimental results obtained up to now. Any contradiction with prediction of the QM may be ignored because of this belief. Such attitude turns QM into no scientific theory, according to the criterion of demarcation between what is and is not genuinely scientific by Karl Popper: a theory should be considered scientific if and only if it is falsifiable. Therefore any experimental results contradicting to the quantum formalism should be at the centre of attention in order QM could be considered as the scientific theory. I would like to draw reader's attention to two of such results. 

According to the quantum formalism (10) two permitted states $n$ and $n+1$ have minimum energy at $\Phi = (n+0.5)\Phi_{0}$. The single-shot measurement should give a result corresponding to the one $n$ or other $n+1$ state with the probability dependent on the $\delta \Phi = \Phi - (n+0.5)\Phi_{0}$ value. The two states $I_{p,n}(\delta \Phi) \propto n - \Phi /\Phi_{0}$ or $I_{p,n+1}(\delta \Phi) \propto n+1 - \Phi /\Phi_{0}$ observed at measurements of the "flux qubit" persistent current $I_{p}(\delta \Phi)$ in \cite{1Shot02} and also for some samples in \cite{1Shot04} corroborate the prediction of the quantum formalism. But besides these results a $\chi $-shaped crossing of the  $I_{p,n}(\delta \Phi)$ and $I_{p,n+1}(\delta \Phi)$ dependencies is observed in \cite{1Shot04}. The authors \cite{1Shot04} interpret this $\chi $-shaped crossing as the single-shot readout of macroscopic quantum superposition of   "flux qubit" states and its absence as classical behavior. The false assumption \cite{QI2008} by the authors \cite{1Shot04} of a possibility to observe the quantum superposition witnesses once again that the QM created by {\it "sleepwalkers"} on the base of {\it the soothing philosophy-or religion-of Heisenberg-Bohr} is misinterpreted by many modern physicists. But the excellent experimental results \cite{1Shot04} are trustworthy. The single-shot readout of the value $I_{p} = 0$ at $\Phi = (n+0.5)\Phi_{0}$, forbidden according to (10), may challenge the quantum formalism. 

Other challenge has been revealed at measurements of magnetic dependencies of the critical current $I_{c+}(\Phi /\Phi_{0})$, $I_{c-}(\Phi /\Phi_{0})$ of asymmetric (with different width $w_{w} > w_{n}$ of half-rings) rings \cite{PCJETP07,JETP07J,PRL06Rej}. According to (10) the critical current of a symmetric ($w_{w} = w_{n}$) ring, measured as it is shown on Fig.1 of \cite{PCJETP07} and \cite{PRL06Rej}, should not depend on the measuring current $I_{ext}$ direction  $I_{c+} = I_{c-} = I_{c}$ and the maximum of the oscillations $ I_{c}(\Phi /\Phi_{0}) = I_{c0} - 2|I_{p}(\Phi /\Phi_{0})|$ should be observed at $\Phi = n\Phi_{0}$ and the minimum at $\Phi = (n+0.5)\Phi_{0}$. The measurements corroborate these predictions, see Fig.2 in \cite{JETP07J}. But measurements of the $I_{c+}(\Phi /\Phi_{0})$, $I_{c-}(\Phi /\Phi_{0})$  dependencies of asymmetric $w_{w} > w_{n}$ rings \cite{PCJETP07,JETP07J,PRL06Rej} challenge the quantum formalism. According to the quantum formalism (10) the critical current anisotropy $I_{c,an}(\Phi /\Phi_{0}) = I_{c+}(\Phi /\Phi_{0}) - I_{c-}(\Phi /\Phi_{0}) $ should appear in the asymmetric ring because of the change of the $I_{c+}(\Phi /\Phi_{0})$, $I_{c-}(\Phi /\Phi_{0})$ functions at $w_{w} > w_{n}$, see Fig.19 \cite{PCJETP07} and  Fig.3 \cite{PRL06Rej}. In contrast to this prediction the measurements have revealed that the asymmetry $I_{c,an}\neq 0$ appears because of changes in the arguments of the functions rather than the functions themselves: $I_{c+}(\Phi /\Phi_{0}) = I_{c-}(\Phi /\Phi_{0})$ at $w_{w}/w_{n} = 1$ and $I_{c+}(\Phi /\Phi_{0} + 0.25) \neq I_{c-}(\Phi /\Phi_{0}- 0.25)$ at $w_{w}/w_{n} \geq  1.25$ \cite{JETP07J}. The shift $\Delta \phi /2 \leq  0.25$ of the $I_{c+}(\Phi /\Phi_{0})$, $I_{c-}(\Phi /\Phi_{0})$  dependencies observed on asymmetric rings \cite{PCJETP07,JETP07J} is quite impossible according to the quantum formalism (10) and contradicts to the resistance dependencies $R(\Phi /\Phi_{0})$ measured on the same rings \cite{JETP07J}. The measurements have revealed also a contradiction concerning observation of the two states $n$ and $n+1$ at $\Phi = (n+0.5)\Phi_{0}$ \cite{PRL06Rej}. The observations of the zero rectified voltage $V_{dc} \propto \overline{v} \propto \overline{n} - \Phi /\Phi_{0} = (0.5 +(-0.5))/2 = 0$ and the maximum of the resistance $\Delta R \propto \overline{v^{2}} \propto \overline{(n - \Phi /\Phi_{0})^{2}} = (0.5^{2} +(-0.5)^{2})/2 $ at $\Phi = (n+0.5)\Phi_{0}$ give evidence of two permitted states with the same minimum energy $\propto v^{2} \propto (n - \Phi /\Phi_{0})^{2} = (1/2)^{2} = (-1/2)^{2}$ \cite{PRL06Rej}. But the two states $n$ and $n+1$ are not observed on the $I_{c+}(\Phi /\Phi_{0})$, $I_{c-}(\Phi /\Phi_{0})$ dependencies measured on the same asymmetric rings \cite{PRL06Rej}.

\subsection{Challenges to the universality of some quantum principles}
%\label{sec:4.3}
Most physicists disregard any challenge to the universality of QM. Although Angelo Bassi and GianCarlo Ghirardi have adduced persuasive arguments against the universal validity of the superposition principle \cite{Ghirardi2000} most authors continue to believe in the universality of this principle and apply it without a moment's hesitation even to macroscopic systems. The lively discussion \cite{AfsharFPh10,AfsharFPh09,AfsharFPh08,AfsharFPh07,AfsharFPh07Af} about the Afshar experiment \cite{AfsharExpSPIE,AfsharExpAIP} demonstrates the belief in the universality of the quantum principles of complementarity and uncertainty. The heated debate implies that these principles should be universally wrong or universally correct. But the complementarity and uncertainty principle introduced by Bohr and Heisenberg in the limits of their positivism point of view should not be valid for description of all quantum phenomena. Einstein confessed that he had been unable to achieve the sharp formulation of Bohr's principle of complementarity {\it "despite much effort which I have expended on it"} \cite{Einstein1949}. Most physicists rather believe in this principle than know its sharp formulation. The debate about the essence of complementarity is in progress  up to now \cite{AndreiCo,Plotnits}. Reading numerous elucidation by Bohr, one may connect this principle with {\it the circumstance that the study of the complementary phenomena demands mutually exclusive experimental arrangements} \cite{Bohr1949}. The same is in the formulation of the Bohr's complementarity principle proposed by  Arkady Plotnitsky: {\it "in considering complementary conjugate variables in question in quantum mechanics (in contrast to those of classical physics) we deal with two mutually exclusive experimental arrangements"} \cite{Plotnits}. 

But it is not correct that experimental arrangements should be mutually exclusive for all conjugate variables and in all cases. The experimental arrangements for measurement of position and momentum are not merely mutually non-exclusive but are the same in the method learned in the primary school. The momentum $p = mv$ of a particle with a mass $m$ is measured $p = m(z_{2}-z_{1})/(t_{2}-t_{1})$ according to this method with help of measurement of the time $t_{1}$ and  $t_{2}$ when the particle passes points $z_{1}$ and $z_{2}$. The particle may be fullerene molecule with mass $m \approx  1.4 \ 10^{-24} \ kg$ used in \cite{Zeilin02} for verification of the Heisenberg uncertainty relation $\Delta x \Delta v_{x} > \hbar /2m$ along the direction $x$ perpendicular to the velocity $v_{z}$. The molecule passing the points $z_{1}$ and $z_{2}$ at the time $t_{1}$ and $t_{2}$ can be detected with help of detection laser used in \cite{Zeilin02}. The velocity value $v_{z} = z/t$ can be measured with the uncertainty $\Delta v_{z} \approx v_{z}(\Delta z/z + \Delta t/t)$ at $z = z_{2} - z_{1} \gg  \Delta z$, $t = t_{2} - t_{1} \gg  \Delta t$. Thus, we can make the product of the velocity $\Delta v_{z}$ and coordinate $\Delta z$ uncertainties $\Delta z  \Delta v_{z} \approx \Delta z  v_{z}( \Delta z/z + \Delta t/t)  $ how any small, contrary to the Heisenberg uncertainty relation $\Delta z \Delta v_{z} > \hbar /2m$, increasing the distance $z = z_{2} - z_{1}$ and the time $t = z/v_{z}$. The uncertainty relation $\Delta z \Delta v_{z} > \hbar /2m \approx 0.3 \ 10^{-10} \ m^{2}/c $ for fullerene molecule with real velocity $v_{z} \approx 100 \ m/c$ \cite{Zeilin02} should be violated at a quite accessible distance $z > 3 \ m$ and accessible measurement inaccuracy of coordinates $\Delta z < 10^{-6} \ m$ and time $\Delta t < 10^{-8} \ c$. 

Thus, the method of the velocity measurement learned in the primary school casts doubt on the universality of both complementarity and uncertainty principle. The Zeilinger's team has confirmed the Heisenberg uncertainty relation for such complex, massive particles as fullerene molecules $C_{70}$ with $m = 840 \ amu$ for the direction perpendicular to their velocity \cite{Zeilin02}. They could also verify this relation for the direction along the fullerene velocity. Violation of the uncertainty relation at the experimental verification will mean that the uncertainty principle is not wrong but is only non-universal. Quantum principles describing only phenomena should not be universal in spite of almost universal belief in their universality. 

\section{Conclusion}
%\label{sec:5}
Richard Feynman stated impartially {\it that no one understands quantum mechanics} \cite{Feynman1967}. Most physicists believe in quantum mechanics. But scientist should understand, at least, what he believes in. Unfortunately the QM, in which the majority believe, differs in essence from the QM created by its founding fathers. The majority decline to understand that the subject of the orthodox QM differs basically from the subject of all other physical theories. The fundamental obscurity in QM should be associated rather with epistemology than with problems of theory or experiment. Einstein accentuated fairly that {\it "Science without epistemology is - insofar as it is thinkable at all - primitive and muddled"} \cite{Einstein1949}. The fundamental obscurity unmasked by Bell and other experts, the numerous conflicting interpretations and the controversy going on during many decades testify against QM as an intelligible scientific theory. QM is muddled just because of the total neglect by most physicists its epistemic problems. The majority opinion prevailed during a long period that the questions "What does QM describe?", "What can we observe in quantum phenomena?" and so on are irrelevant or metaphysical. As the authors of the book \cite{QuCh2006} note fairly {\it "During this period, the wonderful difficulties of quantum mechanics were largely trivialized, swept aside as unimportant philosophical distractions by the bulk of the physics community"}. 

The neglect of the profound questions discussed by the founding fathers of quantum theory has led to regrettable results. These questions do not have universally recognized answers up to now. A realistic theory of quantum phenomena, the necessity of which was vindicated insistently by Einstein, can not be created up to now. It may be such theory is impossible totally. Nevertheless the true comprehension of the well-grounded attacks of Einstein, Bell and others on QM is crucially necessary. QM without such comprehension has become rather scholastic than physical theory. Modern authors make both funny and grandiose mistakes \cite{WakeUp} because of the naive realism inherent in the interpretation of QM by most physicists. Therefore it is needed to explain widely the positivism of the orthodox QM created by Heisenberg, Bohr and others. And it is even more important to understand that this positivism should not be valid universally for the description of all quantum phenomena. Bell tried to explain just this non-universality. According to his opinion {\it the founding fathers were in fact wrong} when they {\it decided even that no concepts could possibly be found which could permit direct description of the quantum world} \cite{Bell1984}. Bell as well as Einstein, in contrast to most physicists, was conscious of inadmissible consequences of the repudiation of realism. His works as well as the criticism of genius against QM by Einstein are misunderstood by most modern authors. Violations of the Bell's inequalities testify against a possibility of realistic description of some quantum phenomena. But it is mistake to jump to conclusions about universality of this impossibility and to use $\psi $ - function for description of all quantum phenomena even macroscopic one. It is important to remember that physics is empirical science. It is needed to revise what quantum phenomena exactly can inevitably threaten our natural aspiration for realism. The unthinking utilization of the superposition principle, contradicting realism, results to both epistemic and practical mistakes, first of all because of the problem of quantum computer as a real device.

\end{document}